\documentclass[twocolumn,secnumarabic,amssymb, nobibnotes, superscriptaddress, nofootinbib, aps, prd]{revtex4-2}

\usepackage{hyperref}       
\usepackage{amsmath}
\numberwithin{equation}{section}
\usepackage{adjustbox}
\usepackage{graphicx}
\usepackage{booktabs}       
\usepackage{amsfonts}       
\usepackage{nicefrac}       
\usepackage{microtype}
\usepackage{cleveref}
\usepackage{orcidlink}
\usepackage{xcolor}
\usepackage{multirow}
\usepackage{todonotes}
\usepackage{makecell}
\usepackage{array}
\usepackage{changes}
\definechangesauthor[name={Prashant}, color=blue]{PT}
\usepackage[normalem]{ulem}

\begin{document}

\title{Effect of Dark matter and $\sigma$-cut potential on radial and non-radial oscillation modes in neutron stars}%

\author{Prashant Thakur \orcidlink{0000-0003-4189-6176}}
\email{p20190072@goa.bits-pilani.ac.in}
\affiliation{Department of Physics, BITS-Pilani, K. K. Birla Goa Campus, Goa 403726, India}
\affiliation{Department of Physics, Yonsei University, Seoul, 03722, South Korea}
\author{Ishfaq Ahmad Rather~\orcidlink{0000-0001-5930-7179}}
\email{rather@astro.uni-frankfurt.de}
\affiliation{Institut f\"{u}r Theoretische Physik, Goethe Universit\"{a}t, 
Max-von-Laue-Str.~1, D-60438 Frankfurt am Main, Germany}
\author{Y. Lim~\orcidlink{0000-0001-9252-823X}} 
\email{ylim@yonsei.ac.kr}
\affiliation{Department of Physics, Yonsei University, Seoul, 03722, South Korea}


\begin{abstract}
 We study the mesonic nonlinear (NL) interaction equation of state (EoS) employing the relativistic mean-field model and investigate the effect of $\sigma$-cut potential (NL-$\sigma$ cut) and dark matter (NL DM) on the non-radial and radial oscillation modes of neutron stars. For NL-$\sigma$ cut, we include the $\sigma$-cut potential $U_{cut} (\sigma)$ to study its effect. For the dark matter, we use the neutron decay anomaly model. For each model, we investigate two extreme EoSs, stiff and soft, that cover the entire allowed parameter range from the given model, consistent with the current astrophysical constraints. The EoS and the stellar properties, such as mass and radius, are calculated, and the effect of $\sigma$-cut and DM is discussed. Both non-radial and radial oscillation modes are computed in the general relativistic framework. We study the non-radial $f$ and $p_1$ mode frequency, damping time, and some qusi-universal relations connecting the frequencies of the $f$-mode to the average density and compactness. The analysis showed that the $f$ and $p_1$ mode frequencies at both 1.4~$M_{\odot}$ and the maximum mass configuration are higher in the NL DM model compared to the NL and NL-$\sigma$ models. The consistent alignment between our prior parameterizations and current calculations strongly supports the existence of quasi-universal relations that hold true irrespective of the particular matter components involved. For the radial oscillations, we compute 10 lowest-order modes ($f$,  $p$), study the radial perturbations as well as the large frequency separation with NL-$\sigma$ cut and NL DM EoS, showing that the microphysics involved in the NS EoS is imprinted on the frequency separation between different nodes.

\end{abstract}
\maketitle

\section{Introduction}

Neutron stars (NSs), the densest observed stars, serve as natural laboratories for studying cold, dense nuclear matter. Their structure and properties are governed by the equation of state (EoS), which remains uncertain due to the nonperturbative nature of strong interactions, especially at densities beyond nuclear saturation ($\rho_0$), where exotic phases may exist. The EoS can be constrained from multiple sources. Radio observations of $\sim 2.0\,M_\odot$ pulsars~\cite{Demorest:2010bx, Antoniadis:2013pzd, Fonseca:2021wxt} require strong repulsion to counteract gravity. Gravitational waves from GW170817~\cite{PhysRevLett.119.161101, LIGOScientific:2017ync, PhysRevLett.121.161101} constrained tidal deformability, ruling out overly stiff EoS models~\cite{Typel2009, PhysRevC.55.540}. X-ray observations by Neutron Star Interior Composition Explorer (NICER)~\cite{Riley:2021pdl, miller2021, Miller_2019a, Riley_2019, Salmi:2024aum, dittmann2024precisemeasurementradiuspsr, Choudhury:2024xbk} further refine constraints on high-density matter.

In recent decades, a wide range of nuclear energy density functional models have been
used to establish a nuclear EoS and investigate the NS properties\,\cite{Dutra:2012mb,lim2018neutron,Lim:2018xne,Lim:2019som,Lim:2023dbk}. 
Among these, relativistic mean-field (RMF) models have been widely employed due to their high degree of parameterizability and their microphysically motivated structure. RMF models are based on a covariant field-theoretical framework that naturally incorporates relativistic effects, which are essential for describing the behavior of dense matter inside neutron stars as well as finite nuclei,\cite{Glendenning:1991es, Horowitz2001, Pais:2016xiu}. Models incorporating effective Lagrangian densities with $\sigma$, $\omega$, $\rho$, and $\delta$ meson couplings to nucleons, along with non-linear self-interactions and cross-coupling terms, have been extensively analyzed,\cite{Lalazissis:1996rd, Long:2003dn, Chen2014, Lim:2017rqk}. These formulations have been applied to both nucleonic and hyperonic matter and benchmarked against constraints from nuclear matter properties and astrophysical observations of NS masses,\cite{Dhiman:2007ck, Pradhan:2022txg}.

Despite this, relativistic mean-field (RMF) models are preferred because they are motivated by their high degree of parameterizability, combined with their
microphysically well-motivated approach. They have been implemented effectively due to their capability in describing matter with relativistic effects, important for dense matter such as in NSs, as well as finite nuclei \cite{Glendenning:1991es, Horowitz2001, Pais:2016xiu}. Among them, RMF models with effective Lagrangian densities consisting of $\sigma$, $\omega$, $\rho$, and $\delta$ mesons couplings with nucleons, including non-linear self-interactions and cross-coupling terms involving these mesons,
have been analyzed \cite{Lalazissis:1996rd, Long:2003dn, Chen2014, Lim:2017rqk}. These models have been
extensively analyzed for both nucleonic and hyperonic matter,
and their predictions have been compared with the constraints
derived from nuclear matter properties and observations of
NS masses in astrophysical environments \cite{Dhiman:2007ck, Pradhan:2022txg}.

Our research explores the RMF framework's non-linear model. Although experimental data provides strong constraints on the EoS of nuclear matter near and below saturation density, its behavior at higher densities is still unknown. Two-solar-mass NS observations indicate that the EoS needs to be sufficiently rigid at high densities. 
To address this requirement,
the $\sigma$ cut potential technique was presented in \cite{Maslov:2015lma} in order to account for this stiffening beyond roughly twice the saturation density. This technique efficiently stiffens the EoS while maintaining nuclear matter properties close to $\rho_0$ \cite{Ma:2022fmu} by enforcing a significant increase in the mean-field self-interaction potential above the nuclear saturation density. An additional factor influencing the EoS is the possible presence of dark matter (DM) in NSs. DM constitutes approximately 24\% of the total matter in the universe, yet its fundamental nature remains one of the greatest unresolved questions in modern astrophysics. Although its existence has been inferred through gravitational interactions on both cosmological and galactic scales, direct detection remains challenging. Investigating the role of DM in compact astrophysical objects like NSs provides a unique opportunity to explore its properties and potential interactions with baryonic matter.

Several studies have estimated the DM fraction in NSs based on observational constraints. For instance, \citet{Ivanytskyi:2019wxd} reported that the DM fraction is approximately $f_{\chi}^* = 1.6 \pm 0.4\%$ near PSR J0348+0432 and $f_{\chi}^* = 1.35 \pm 0.35\%$ around PSR J0740+6620. Additional constraints from sub-GeV bosonic dark matter models, as discussed in Ref.~\cite{Karkevandi:2021ygv}, indicate that current observational data favor low DM fractions, typically below 5\%. Meanwhile, \citet{Rutherford:2022xeb} propose an upper limit of $f_{\chi} \leq 20\%$. Furthermore, \citet{Ciarcelluti:2010ji} suggests that DM can contribute to the formation of highly compact neutron stars and may explain the existence of massive pulsars, such as PSR J1614-2230 \cite{2010Natur.467.1081D} and PSR J0348+0432 \cite{Lynch_2013}. Their study indicates that an NS can reach a mass of $2\,M_{\odot}$ with a DM fraction of 15\%, while a fraction of 70\% can yield a mass of $1.8\,M_{\odot}$. Additionally, \citet{Goldman:2013qla} suggests that an NS with a DM fraction of 50\% can achieve a mass of $2\,M_{\odot}$. \citet{Pitz:2024xvh} showed that a DM fraction of around 90\% is required to explain the unusual measurement of XTE J1814-338 \cite{Kini:2024ggu}. Recent studies have extended this understanding by incorporating the effects of DM within NSs \cite{ Nelson:2018xtr, Das:2020ecp, Karkevandi:2021ygv, Giangrandi:2022wht, Rutherford:2022xeb, Shakeri:2022dwg, Singh:2022wvw, Thakur:2023aqm} which significantly alter the EoS, potentially leading to observable changes in NS characteristics. 

Two widely studied approaches for investigating neutron stars containing dark matter are well documented in the existing literature.
One approach considers non-gravitational interactions, using mechanisms like the Higgs portal \cite{Dutra:2022mxl, Lenzi:2022ypb, Hong:2024sey, Das:2018frc, Flores:2024hts, Das:2020vng}. The other approach considers only gravitational interactions, treated as a two-fluid system \cite{Collier:2022cpr, Miao:2022rqj, Hong:2024sey, Karkevandi:2021ygv, Ruter:2023uzc, Ivanytskyi:2019wxd, Buras-Stubbs:2024don, Rutherford:2022xeb, Hajkarim:2024ecp,Issifu:2024htq}. Another method involves the neutron decay into dark matter particles that accounts for the neutron decay anomaly \cite{Bastero-Gil:2024kjo, PhysRevLett.121.061801, Shirke:2023ktu,Motta:2018bil, Shirke:2024ymc}.

The impact of DM on neutron stars can be further explored through multimessenger astronomy, which combines electromagnetic waves, gravitational waves, neutrinos, and cosmic rays to probe NS mergers and dense matter. With increasing gravitational wave detections, NS asteroseismology has become a key tool for probing the dense matter EoS. \citet{Andersson:1997rn} pioneered gravitational wave asteroseismology by proposing the first asteroseismological relations for neutron star oscillation modes. Later studies introduced similar quasi-universal relations for the $f$-mode frequency and damping time \cite{Lau:2009bu, Chirenti:2015dda, Sotani:2020bey}.
The upcoming ground-based gravitational wave detectors, including the Cosmic Explorer~\cite{Evans:2021gyd} and the Einstein Telescope~\cite{Punturo_2010, 2021arXiv211106990K, Abac:2025saz}, are anticipated to enhance sensitivity by an order of magnitude. This advancement will enable the detection of these oscillations, as these detectors will achieve unprecedented sensitivity in the frequency range of approximately $10$ Hz to a few kHz.

Detection of oscillation spectra provides key information on NS mass, frequency, tidal Love numbers, damping times, and moments of inertia, helping to constrain the EoS of dense matter~\cite{Providencia:2023rxc}. Newly formed NSs in supernovae oscillate at characteristic frequencies determined by the restoring force, which may have different origins~\cite{PhysRevLett.116.181101, PhysRevLett.108.011102, Chirenti_2017}. These oscillations are classified as radial and non-radial modes. While radial modes do not directly produce gravitational waves (GWs), their coupling with non-radial modes enhances GW emission, improving detectability~\cite{PhysRevD.73.084010, PhysRevD.75.084038}. \citet{Chirenti_2019} noted that BNS post-merger remnants can form hyper-massive NSs, influencing gamma-ray bursts and radial oscillations. Their high-frequency modes ($\sim 1$–$4$ kHz) may be detectable by LIGO, Virgo, and KAGRA~\cite{PhysRevLett.122.061104, 10.1093/ptep/ptac073}, with improved sensitivity in third-generation detectors~\cite{Evans:2021gyd, Punturo_2010, 2021arXiv211106990K}.

Different families of oscillation modes arise from distinct physical conditions~\cite{kokkostas}. Radial modes are classified into two types: one from the high-density core and the other from the low-density envelope, separated by a "wall" in the adiabatic index due to the neutron drip transition~\cite{1997A&A...325..217G}. This effect is universal across realistic EoSs and could serve as a unique probe of NS interiors via future GW observations. Extensive studies have explored radial oscillations, including exotic phases like dark matter and deconfined quark matter~\cite{Rather:2023dom, Rather:2024hmo, 10.1093/mnras/stac2622, kokkostas, Sen:2022kva, PhysRevD.101.063025, PhysRevD.98.083001, Routaray:2022utr}.

Non-radial oscillations in NSs include $f$, $g$, and $p$ modes (polar) as well as $r$ and $w$ modes (axial)~\cite{Kokkotas:1999bd, Benhar:2004xg}. However, $w$-modes, with frequencies of 5–12 kHz, are typically not excited during NS mergers. In contrast, the $f$-mode plays a crucial role in GW detection, as it emits strong signals within the 1–3 kHz range, making it observable by current and future detectors. The $f$-mode frequency is linked to NS properties, including tidal deformability in mergers, compactness~\cite{Andersson:1997rn}, moment of inertia~\cite{Lau:2009bu}, and static tidal polarizability~\cite{PhysRevD.90.124023}. It can be excited by various astrophysical events such as NS formation~\cite{Ferrari:2002ut}, starquakes~\cite{Mock_1998, Kokkotas:1999mn}, and magnetar activity~\cite{LIGOScientific:2019ccu, LIGOScientific:2022sts}. For GW170817, the $f$-mode frequency was estimated in the range of 1.43–3.18 kHz~\cite{Kunjipurayil:2022zah}.

The interior composition and structure of NSs remain among the most profound enigmas in modern astrophysics, offering unique insights into fundamental physics under extreme conditions unreproducible in terrestrial laboratories. Our work builds on ~\citet{Thakur:2024btu}, where a Bayesian framework was used to compare the NL, NL-$\sigma$, and NL DM models by incorporating constraints from nuclear experiments and astrophysical observations. The study found that, excluding PREX-II data, the NL-$\sigma$ model showed the highest Bayesian evidence, highlighting its better consistency with observational constraints. To further bridge the gap between theoretical predictions and observational constraints on NS properties, we extend this analysis by investigating radial and non-radial oscillation modes in all three models. The oscillation modes of NSs—both radial and non-radial—serve as powerful probes of their internal structure, carrying distinctive signatures of the underlying EoS. By investigating how modifications to the RMF model by including $\sigma$-cut potential and DM affect these oscillation patterns, we aim to provide observationally testable predictions that can help discriminate between competing theoretical frameworks. This work is particularly timely given the recent advancements in GW astronomy and x-ray timing observations, which have begun to place increasingly stringent constraints on NS properties. Furthermore, understanding the role of exotic components such as dark matter in NS interiors may provide crucial insights into fundamental particle physics, while the quasi-universal relations we explore between oscillation frequencies and stellar parameters could serve as robust diagnostic tools for future GW detections, potentially revealing the imprint of microphysics on macroscopic observables.

 This work is organized as follows: In Section \ref{theory} we explain the RMF model with the addition of $\sigma$-cut potential and DM. Section \ref{oscillation} discusses the formalism for non-radial and radial oscillations. We show the results for non-radial and radial frequencies along with the EoS and TOV solutions in Section \ref{results}. Finally, we discuss our findings in Section \ref{results}.

\section{Theoretical framework and formalism}
\label{theory}

\subsection{NL model}
In the RMF model, the EoS for nuclear matter is described by the interaction of the scalar-isoscalar $\sigma$, the vector-isoscalar $\omega$, and the vector-isovector $\varrho$ mesons. The Lagrangian density is given by \citep{Fattoyev:2010mx,Dutra:2014qga,Malik:2023mnx}
        \begin{equation} \label{lag}
          \mathcal{L}=   \mathcal{L}_N+ \mathcal{L}_M + \mathcal{L}_{NL} +\mathcal{L}_{leptons} ,
        \end{equation} 
with
\begin{equation*}
  \mathcal{L}_{N} = \bar{\Psi}\Big[\gamma^{\mu}\left(i \partial_{\mu}-g_{\omega} \omega_\mu - \frac{1}{2}g_{\varrho} {\boldsymbol{\tau}} \cdot \boldsymbol{\varrho}_{\mu}\right) - \left(m_N - g_{\sigma} \sigma\right)\Big] \Psi ,   
\end{equation*}
representing the Dirac Lagrangian density for the neutron and proton doublet with a bare mass $m_N$, where $\Psi$ denotes a Dirac spinor, $\gamma^\mu$ are the Dirac matrices, and $\boldsymbol{\tau}$ symbolizes the isospin operator.
$\mathcal{L}_{M}$ is the Lagrangian density for the mesons, given by
\begin{eqnarray}
\mathcal{L}_{M}  &=& \frac{1}{2}\left[\partial_{\mu} \sigma \partial^{\mu} \sigma-m_{\sigma}^{2} \sigma^{2} \right] - \frac{1}{4} F_{\mu \nu}^{(\omega)} F^{(\omega) \mu \nu} + \frac{1}{2}m_{\omega}^{2} \omega_{\mu} \omega^{\mu}   \nonumber \\
  &-& \frac{1}{4} \boldsymbol{F}_{\mu \nu}^{(\varrho)} \cdot \boldsymbol{F}^{(\varrho) \mu \nu} + \frac{1}{2} m_{\varrho}^{2} \boldsymbol{\varrho}_{\mu} \cdot \boldsymbol{\varrho}^{\mu} \nonumber ,
\end{eqnarray}
where $F^{(\omega, \varrho)\mu \nu} = \partial^ \mu A^{(\omega, \varrho)\nu} -\partial^ \nu A^{(\omega, \varrho) \mu}$ are the vector meson tensors, and
\begin{eqnarray}
\mathcal{L}_{NL}&=&-\frac{1}{3} b~m_N~ g_\sigma^3 (\sigma)^{3}-\frac{1}{4} c (g_\sigma \sigma)^{4}+\frac{\xi}{4!} g_{\omega}^4 (\omega_{\mu}\omega^{\mu})^{2}  \nonumber \\
&+&\Lambda_{\omega}g_{\varrho}^{2}\boldsymbol{\varrho}_{\mu} \cdot \boldsymbol{\varrho}^{\mu} g_{\omega}^{2}\omega_{\mu}\omega^{\mu} \nonumber ,
\end{eqnarray}
includes the non-linear mesonic terms characterized by the parameters $b$, $c$, $\xi$, and $\Lambda_{\omega}$, which manage the high-density properties of nuclear matter. The coefficients $g_i$ represent the couplings between the nucleons and the meson fields $i = \sigma, \omega, \varrho$, with the masses denoted by $m_i$. The parameter values employed in this model are listed in Table \ref{tab:nl_models}.

Finally, the dynamics of leptons are described by the Lagrangian density  
\begin{equation*}
    \mathcal{L}_{leptons}= \bar{\Psi_l}\Big[\gamma^{\mu}\left(i \partial_{\mu}  
-m_l \right)\Big]\Psi_l ,
\end{equation*} 
where $\Psi_l~(l= e^-, \mu^-)$ denotes the lepton spinor for electrons and muons; 
leptons are treated as non-interacting particles with nucleons and dark matter.

The energy density of baryons and leptons is given by the following expression:
\begin{equation}
\begin{aligned}
\mathcal{E} &= \sum_{i=n,p,e,\mu}\frac{1}{\pi^2}\int_0^{k_{Fi}} \sqrt{k^2+{m_i^*}^2}\, k^2\, dk \\
&+ \frac{1}{2}m_{\sigma}^{2}{\sigma}^{2}+\frac{1}{2}m_{\omega}^{2}{\omega}^{2}+\frac{1}{2}m_{\varrho}^{2}{\varrho}^{2}\\
&+ \frac{b}{3} m_{\rm N} (g_{\sigma}{\sigma})^{3}+\frac{c}{4}(g_{\sigma}{\sigma})^{4}+\frac{\xi}{8}(g_{\omega}{\omega})^{4} + \Lambda_{\omega}(g_{\varrho}g_{\omega}{\varrho}{\omega})^{2},
\end{aligned}
\end{equation}
where $m_i^*=m_i-g_{\sigma} \sigma$ is the effective mass for protons and neutrons. For leptons,  $m_i^*=m_i$. $k_{Fi}$ is the Fermi moment of the particle $i$.
{The $\sigma$, $\omega$ and $\varrho$ are the  mean-field values of the corresponding mesons \cite{Malik:2023mnx}}. 

Once the energy density is obtained for a given EoS model, we can compute the chemical potential of neutrons ($\mu_n$) and protons ($\mu_p$). The chemical potential of electron ($\mu_e$) and muon ($\mu_\mu$) can be computed using the $\beta$-equilibrium condition of: $\mu_n-\mu_p=\mu_e$ and $\mu_e$ = $\mu_\mu$ as well as the charge neutrality: $\rho_p =\rho_e + \rho_\mu$, where $\rho_e$ and $\rho_\mu$ are the electron and muon number density, respectively.

The pressure $P$ is then determined from the following thermodynamic relation:
\begin{equation}
P = \sum_{i}\mu_{i}\rho_{i}-\mathcal{E}.
\label{eq_9}
\end{equation}
which can be expressed as: 

\begin{equation}
\begin{aligned}
P &= \sum_{i=n,p,e,\mu}\frac{1}{\pi^2}\int_0^{k_{Fi}} \sqrt{k^2+{m_i^*}^2}\, k^2\, dk \\
&- \frac{1}{2}m_{\sigma}^{2}{\sigma}^{2}-\frac{1}{2}m_{\omega}^{2}{\omega}^{2}-\frac{1}{2}m_{\varrho}^{2}{\varrho}^{2}\\
&- \frac{b}{3}(g_{\sigma}{\sigma})^{3}-\frac{c}{4}(g_{\sigma}{\sigma})^{4}+\frac{\xi}{8}(g_{\omega}{\omega})^{4} + \Lambda_{\omega}(g_{\varrho}g_{\omega}{\varrho}{\omega})^{2},
\end{aligned}
\end{equation}

\subsection{NL-$\sigma$ Cut}
The inclusion of the $\sigma$-cut potential $U_{\text{cut}}(\sigma)$ in the RMF Lagrangian, as described in \cite{Maslov:2015lma, Zhang:2018lpl, Thakur:2024scc}, is implemented for further exploration in this work.
According to \cite{Maslov:2015lma}, the $U_{\text{cut}}(\sigma)$ has a logarithmic form and only impacts the $\sigma$ field at high density. It is determined by,

\begin{eqnarray}
U_{cut}(\sigma) = \alpha \ln [ 1 + \exp\{\beta(g_{\sigma}\sigma/m_{N}-f_{s})\}] ,
\end{eqnarray}
where $\alpha$ = $m_{\pi}^{4}$ and $\beta$ = 120 \cite{Maslov:2015lma}. values of $f_s$ are taken from our previous study\cite{Thakur:2024btu} and are mentioned in Table \ref{tab:nl_models}.

The strategy is to quench the growth of the scalar field at
densities larger than saturation density $\rho_0$.

The Lagrangian density for $\mathcal{L}_{NL-\sigma \rm cut}$  with $U_{cut}(\sigma)$ is given as,
\begin{eqnarray}
    \mathcal{L}_{NL-\sigma cut}&=&\mathcal{L}_{NL} - U_{cut}(\sigma) ,
\end{eqnarray}
and the  meson fields are determined from the  equations 
		\begin{eqnarray}
			{\sigma}&=& \frac{g_{\sigma}}{m_{\sigma,{\rm eff}}^{2}}\sum_{i} \rho^s_i\label{sigma} ,\nonumber \\
			{\omega} &=&\frac{g_{\omega}}{m_{\omega,{\rm eff}}^{2}} \sum_{i} \rho_i \label{omega} , \nonumber \\
			{\varrho} &=&\frac{g_{\varrho}}{m_{\varrho,{\rm eff}}^{2}}\sum_{i} I_{3} \rho_i, \label{rho}
		\end{eqnarray}
 where $\rho^s_i$ and $\rho_i$ are, respectively, the scalar density and the number density of nucleon $i$. The effective masses for mesons are given by
 \begin{eqnarray}
     m_{\sigma,{\rm eff}}^{2}&=& m_{\sigma}^{2}+{ b ~m_N ~g_\sigma^3}{\sigma}+{c g_\sigma^4}{\sigma}^{2} + \frac{U_{cut}^{'}(\sigma)}{\sigma} , \nonumber \\
    m_{\omega,{\rm eff}}^{2}&=& m_{\omega}^{2}+ \frac{\xi}{3!}g_{\omega}^{4}{\omega}^{2} +2\Lambda_{\omega}g_{\varrho}^{2}g_{\omega}^{2}{\varrho}^{2} , \nonumber \\
    m_{\varrho,{\rm eff}}^{2}&=&m_{\varrho}^{2}+2\Lambda_{\omega}g_{\omega}^{2}g_{\varrho}^{2}{\omega}^{2} , \label{mw}
 \end{eqnarray}
where $U_{cut}^{'}(\sigma)$ is the derivative of  $U_{cut}(\sigma)$ with respect to $\sigma$.
In these equations, the meson fields should be interpreted as their expectation values. The energy density and pressure for this case are \cite{Maslov:2015lma} 
\begin{eqnarray}
   \mathcal{E}_{\sigma \rm cut} &=  \mathcal{E} +U_{cut}(\sigma) ,  \\
   P_{\sigma \rm cut} &=  P - U_{cut}(\sigma).
\end{eqnarray}

\subsection{Dark Matter}
Current interpretation of neutron decay experimental data points to the possibility of events outside the realm of the standard model of physics~\cite{PhysRevLett.120.202002,PhysRevLett.120.191801,PhysRevLett.121.061801}. Neutrons mostly experience $\beta$-decay:
\begin{equation}
n \rightarrow p + e^- + \bar{\nu}_e.
\end{equation}
The two neutron lifetime experiments—the bottle experiment and the beam experiment—produce two distinct neutron lifetimes measurements as $\tau_{\rm bottle}=879.6\pm 0.6$ s~\cite{Mampe:1993an,Serebrov:2004zf,Pichlmaier:2010zz,Steyerl:2012zz,Arzumanov:2015tea,Pattie:2017vsj} and $\tau_{\rm beam}=888.0\pm 2.0$ s~\cite{Yue:2013qrc, Byrne:1996zz}, respectively. 
A reconciliation of our understanding of fundamental interactions is necessary, as the two neutron lifetime measurements differ by $4\sigma$.
An interesting hypothesis put out by the authors in Ref.~\cite{PhysRevLett.120.191801} is that the anomaly in the neutron lifetime measurement may be explained by novel neutron decay channels into dark matter particles. 

In NS physics, these new decay channels—where neutrons decay into dark matter particles—may be of interest. NSs being neutron-rich can be the perfect test sites for the predicted decay of neutrons into dark matter particles, according to a number of research studies carried out recently~\cite{Husain:2022bxl, Bastero-Gil:2024kjo, PhysRevLett.121.061801, Shirke:2023ktu, Motta:2018bil, Shirke:2024ymc}.

In this work, we examine the effect of neutron decay on neutron star dynamics using the decay channel involving baryon number violation beyond the standard model (BSM) interaction,
\begin{equation}\label{eqn:darkdecay}
n\rightarrow\chi+\phi~,
\end{equation}
where $\chi$ is a dark spin-1/2 fermion, and $\phi$ is a light dark boson. Other decay channels of neutrons are also possible, e.g., $n\rightarrow \chi+\gamma$, and $n\rightarrow \chi+e^+e^-$.
However, not all decay channels are favored phenomenologically;
 e.g., laboratory experiment puts stringent constraints on the decay channel $n\rightarrow \chi+\gamma$~\cite{Tang:2018eln}. The decay channel $n \rightarrow \chi + \phi$ is especially intriguing in the context of NS physics, as it can be argued that the light dark matter boson $\phi$ would quickly escape a NS, rendering it insignificant. Conversely, some of 
neutrons within an NS will transform into fermionic DM $\chi$ due to the BSM interaction. Physically, these DM particles will experience the gravitational potential of a NS and will reach thermal equilibrium with the surrounding NS matter. This sets the equilibrium condition 
\begin{equation}\label{eqn:chemicalequilibrium}
    \mu_{\chi}=\mu_{n}~.
\end{equation}

Nuclear stability requires $ 937.993 \, \text{MeV} < m_{\chi} + m_{\phi} < m_n = 939.565 \, \text{MeV} $ \cite{Motta:2018bil,Shirke:2023ktu}. For the dark particles to remain stable and avoid further beta decay, the condition $ |m_{\chi} - m_{\phi}| < m_p + m_e = 938.783 \, \text{MeV} $ must be met \cite{Fornal:2020gto}. In this work, we set $m_\chi=938.0$ MeV. 

To account for DM self-interactions, we introduce vector interactions between dark particles, described by
\begin{equation}
    \mathcal{L} \supset -g_V \Bar{\chi} \gamma^{\mu} \chi V_{\mu} - \frac{1}{4} V_{\mu\nu} V^{\mu\nu} + \frac{1}{2} m_V^2 V_{\mu} V^{\mu}~,
\end{equation}
where $ g_V $ is the coupling strength between the fermionic dark matter and the vector meson,
and $ m_V $ is the mass of the vector boson. This introduces an additional interaction term in the energy density, beyond the free fermion part. The energy density of DM is given by
\begin{equation}
    \mathcal{E}_{DM} = \frac{1}{\pi^2} \int_{0}^{k_{F_{\chi}}} k^2 \sqrt{k^2 + m_{\chi}^2} \, dk + \frac{1}{2} G_{\chi} n_{\chi}^2,
    \label{eqn:endens_dm}
\end{equation}
where,
    $G_{\chi} = \left( {g_V}/{m_V} \right)^2$,  $n_{\chi} = {k_{F_{\chi}}^3}/{3 \pi^2}$,
and
$\mu_\chi = \sqrt{k^2 + m_\chi^2} + G_{\chi} n_\chi $.

Furthermore, using the thermodynamic relation, the pressure is given by
\begin{equation}
P_{tot} = \sum_{i}\mu_{i}\rho_{i}-\mathcal{E}_{tot}.
\label{eq_9}
\end{equation}

This study examines a scenario where a chemical equilibrium is established between ordinary matter and the dark sector. The DM population vanishes as the neutron density approaches zero. Thus, we used a single-fluid TOV approach \cite{Shirke:2023ktu, Shirke:2024ymc} instead of the two-fluid method in \cite{Husain:2022bxl}. We add the DM energy density ($\mathcal{E}_{DM}$) to the hadronic matter energy density ($\mathcal{E}$) to get the total energy density ($\mathcal{E}_{\rm tot} = \mathcal{E} + \mathcal{E}_{DM}$). The pressure is calculated using Eq. (\ref{eq_9}).
The prior range for the dark matter self-interaction strength $G_\chi$ is initially motivated by the theoretical analysis presented in Ref.~\cite{Shirke:2023ktu}. A lower limit of $G_\chi > 1.6\,\text{fm}^2$ is required to satisfy the tidal deformability constraint $\Lambda_{1.4M_{\odot}} > 70$~\cite{LIGOScientific:2018cki}. An upper bound from $\Lambda$ is not well-defined, as its value becomes insensitive to large $G_\chi$. However, constraints from the core-cusp problem on galactic scales impose an upper limit of $G_\chi \leq 943\,\text{fm}^2$ for a dark matter particle mass of $m_\chi = 938$~MeV. In the present study, we have taken values mentioned in Table \ref{tab:nl_models} from our previous work~\cite{Thakur:2024btu} where a detailed Bayesian inference analysis of the model was carried out.

\begin{table}[h]
    \centering
    \caption{Parameter values for different models considered in this work. Specifically, $B$ and $C$ are $b \times 10^3$ and $c \times 10^3$, respectively. The parameter $f_{s}$ in the NL-$\sigma$ cut model is dimensionless, while the parameter $G_{\chi}$ in the NL DM model is measured in units of ${\rm fm}^2$.}
\begin{tabular}{ c |c c |c c |c c| }
\toprule
\multirow{2}{*}{~Model~}
 & \multicolumn{2}{c}{NL} & \multicolumn{2}{ c}{NL-$\sigma$ cut} & \multicolumn{2}{ c |}{NL DM} \\
\cline{2-7}
                   & ~Soft~     & ~Stiff~    & ~Soft~    & ~Stiff~   & ~Soft~    & ~Stiff~ \\
\hline
$g_\sigma$         &  $8.92$    &  $8.67$    & $8.40$    &  $8.44$   &  $8.30$   & $8.56$ \\
$g_\omega$         &  $10.76$   &  $10.35$   & $9.79$    &  $9.87$   &  $9.57$   & $10.14$\\
$g_\rho$           &  $10.23$   &  $10.22$   & $10.60$   &  $10.18$  &  $10.77$  & $10.20$\\
$B$                &  $4.24$    &  $4.65$    & $5.64$    &  $5.47$   &  $6.10$   & $4.86$\\
$C$                &  $-4.77$   &  $-4.86$   & $-4.56$   &  $-4.75$   &  $-4.77$ & $-4.70$\\
$\xi$              &  ~$0.00074$ &  ~$0.00028$ & ~$0.00063$ &  ~$0.0029$ &  ~$0.0012$ & ~$0.0003$\\
$\Lambda_\omega$   &  $0.038$   &  $0.041$   & $0.072$   &  $0.049$  &  $0.099$  & $0.047$\\
$f_s$              &  $0.0$     &  $0.0$     & $0.57$    &  $0.50$   &  $0.0$    & $0.0$\\
$G_\chi$           &  $-$       &  $-$       & $-$       &  $-$      &  $420$    & $767$\\
\bottomrule
\end{tabular}
\label{tab:nl_models}
\end{table}

\section{Oscillation modes}
\label{oscillation}

 \subsection{Non-radial}
\label{sec:GRR}
In our previous study \cite{Rather:2024nry}, we showed how the $f$-mode frequencies, calculated using both the Cowling approximation and the general relativistic (GR) frameworks, reveal a significant overestimation by the Cowling method of about 10–30\%, highlighting the importance of GR calculations for accurate modeling. Here we employ the full GR formalism, and
to determine the frequencies of the $f$-modes, we solve Einstein's field equations assuming that the gravitational waves represent perturbations to the static background spacetime metric of a non-rotating NS. The perturbed metric is given by
\begin{equation}
g_{\mu \nu} = g^{0}_{\mu \nu}+h_{\mu \nu},
\end{equation}
Only even-parity perturbations of the Regge-Wheeler metric are significant in this context~\cite{Zerilli:1970se}.
A small perturbation, $ h_{\mu \nu} $, is introduced to a static, spherically symmetric background metric, which is described as:
\begin{align}
ds^2 ={}& - e^{\nu(r)} [1+r^l H_0(r)e^{i\omega t} Y_{lm}(\phi,\theta)]c^2 dt^2\nonumber\\&
+e^{\lambda(r)} [1-r^l H_0(r)e^{i\omega t} Y_{lm}(\phi,\theta)]dr^2 \nonumber\\&
+ [1-r^l K(r)e^{i\omega t}Y_{lm}(\phi,\theta)]r^2 d\Omega^2\nonumber\\&
-2i\omega r^{l+1}H_1(r)e^{i\omega t}Y_{lm}(\phi,\theta) dt~dr,
\end{align}
where $ H_0 $, $ H_1 $, and $ K $ represent the radial perturbations of the metric, while the angular dependence is captured by the spherical harmonics $ Y_l^m $. The time dependence of the perturbed metric components can be expressed using the factor $ e^{i\omega t} $ for a wave mode. Here $ \omega $ is a complex quantity, as the waves decay due to the imposed open boundary conditions. The real part of $\omega$ represents the oscillation frequency, while the imaginary part corresponds to the inverse of the wave mode's gravitational wave damping time (positive).

The perturbations of the energy-momentum tensor of the fluid must also be considered in the Einstein equations. The components of the Lagrangian displacement vector $ \xi^a(r, \theta, \phi) $ describe the perturbations of the fluid within the star:
\begin{align}
\xi^r ={}& r^{l-1}e^{-\frac{\lambda}{2}}W Y^l_m e^{i\omega t} ,\label{eq:xi_radial} \nonumber \\
\xi^\theta ={}& -r^{l-2} V \partial_\theta Y_m^l e^{i\omega t}, \nonumber  \\
\xi^\phi ={}& -\frac{r^{l-2}}{ \sin^{2}\theta} V\partial_\phi Y_m^l e^{i\omega t} .
\end{align}
Here, $W$ and $V$ are functions of $r$ that represent fluid perturbations confined to the star's interior.

The gravitational wave equations can then be written as a set of four coupled linear differential equations for the four perturbation functions, $H_1$, $K$, $W$, and $X$, which do not diverge inside a star for any given value of $\omega$~\cite{Lindblom:1983ps}
\begin{align}
r\frac{dH_1}{dr}&=-[l+1+2b e^\lambda+4\pi r^2 e^\lambda(P-\mathcal{E})] H_1
\nonumber\\
&+ e^\lambda[H_0+K-16\pi(\mathcal{E}+P)V]\,, 
\label{eq:ODE_DL1} \\
r\frac{dK}{dr}&= H_0+(n_l+1)H_1 
\nonumber \\
&+[e^\lambda \textrm{Q}-l-1]K-8\pi(\mathcal{E}+P)e^{\lambda/2}W \,, 
\label{eq:ODE_DL2}\\
r\frac{dW}{dr}&=-(l+1)[W+le^\frac{\lambda}{2}V] 
\nonumber \\
&+r^2 e^{\lambda/2}\left[\frac{e^{-\nu/2}X}{ (\mathcal{E}+P) c_{\rm ad}^2}+\frac{H_0}{2}+K\right], 
\label{eq:ODE_DL3}
\end{align}
\begin{align}
&r\frac{dX}{dr}= -lX+\frac{(\mathcal{E}+P)e^{\nu/2}}{2}\times
\nonumber \\
&\times\Bigg\{ (1-e^\lambda \textrm{Q})H_0+(r^2\omega^2e^{-\nu}+n_l+1)H_1 
\nonumber  \\
&+(3e^\lambda\textrm{Q}-1)K
-\frac{4(n_l+1)e^\lambda\textrm{Q}}{r^2}V 
\nonumber\\
&-2\left[\omega^2 e^{\lambda/2-\nu} +4\pi(\mathcal{E}+P)e^{\lambda/2} -r^2\frac{d}{dr} \left(\frac{e^{{\lambda/2}} \textrm{Q}}{r^3} \right)\right]W \Bigg\} \,, \label{eq:ODE_DL4}
\end{align}
where $c_{\rm ad}^2$ is the adiabatic speed of sound for NS matter under oscillations. In this work, we approximate this speed of sound with the equilibrium sound speed $c_{\rm eq}^2=dP/d\mathcal{E}$. 

Perturbations at the center of the star $r=0$ are subject to
the boundary conditions  $X(R) =0$, $W(0) =1$, and
\begin{align}
X(0) ={}& (\mathcal{E}_0+P_0)e^{\nu_0/2} \nonumber \\
& \bigg\{ \left[ \frac{4\pi}{3}(\mathcal{E}_0+3P_0) 
- \frac{\omega^2}{l} e^{-\nu_0} \right]W(0)+\frac{K(0)}{2} \bigg\}, \\
H_1(0) ={}& \frac{lK(0)+8\pi(\mathcal{E}_0+P_0)W(0)}{n_l+1}.
\label{eq:boundary_conditions}
\end{align}
The final boundary condition is derived by solving two trial solutions with 
$ K(0) = \pm(\varepsilon_0+p_0) $ and then forming a linear combination to satisfy the condition 
$ X(r=R) = 0 $, which ensures there are no pressure variations at the surface. By design, 
$ H_0(0) = K(0) $.

At the star's surface, small arbitrary values are assigned to the functions 
$ H_1 $, $ K $, and $ W $, and backward integration is performed until reaching the point where forward 
integration from the star’s center ends. The forward and backward solutions are then matched at this point. 
The quasinormal mode frequency for the star is determined by solving the Zerilli equation,
\begin{equation}
\frac{d^2Z}{dr^{*2}}=[V_Z(r)-\omega^2]Z \,. \label{eq:zerilli}
\end{equation}
The Zerilli function, as expressed in Eq.~(20) of~\cite{Kunjipurayil:2022zah}, depends solely on the perturbation variables $H_1$ and $K$, since the fluid perturbations $W$, $V$, and $X$ vanish outside the star. The value $Z(r)$ at the star's surface is determined using the values of $H_1$ and $K$ at the surface. Beyond the star, Eq.~(\ref{eq:zerilli}) is numerically integrated starting from the surface and extending outward to a distance corresponding to $r = 25~\omega^{-1}$~\cite{Kunjipurayil:2022zah}. The value of $Z$ at $r = 25~\omega^{-1}$ is matched with the corresponding value obtained from the asymptotic expansion of $Z$, which is valid far from the neutron star's surface. To account for the imaginary component of $\omega$, which is over a thousand times smaller than its real counterpart, it is essential to maintain a relative error of $10^{-6}$ in our ODE solver for the variables $H_1$, $K$, $W$, $X$, and $Z$.

\subsection{Radial}
\label{radial}

In a spherically symmetric system with radial motion, the Schwarzschild metric becomes time-dependent, allowing the use of the Einstein field equations to analyze radial oscillation characteristics of a static equilibrium configuration \cite{1966ApJ...145..505B}. Considering the radial displacement $\Delta r$ and pressure perturbation $\Delta P$, the equations for the dimensionless variables $\xi = \Delta r / r$ and $\eta = \Delta P / P$ are perturbed \cite{1977ApJ...217..799C, 1997A&A...325..217G}

 \begin{equation}\label{ksi}
     \frac{d\xi}{dr} = -\frac{1}{r} \Biggl( 3\xi +\frac{\eta}{\gamma}\Biggr) -\frac{P'(r)}{P+\mathcal{E}} \xi(r),
 \end{equation}
 \begin{equation}\label{eta}
 \begin{split}
          \frac{d\eta}{dr} = \xi \Biggl[ \omega^{2} r (1+\mathcal{E}/P) e^{\lambda - \nu } -\frac{4P'(r)}{P} -8\pi (P+\mathcal{E}) re^{\lambda} \\
     +  \frac{r(P'(r))^{2}}{P(P+\mathcal{E})}\Biggr] + \eta \Biggl[ -\frac{\mathcal{E}P'(r)}{P(P+\mathcal{E})} -4\pi (P+\mathcal{E}) re^{\lambda}\Biggr] ,
      \end{split}
 \end{equation}
 where $\omega$ is the frequency oscillation mode and $\gamma$ is the adiabatic relativistic index defined in terms of energy density ($\mathcal{E}$), pressure ($P$), and speed of sound ($c_s^2/c^2 = dP/d\mathcal{E}$) as
 \begin{eqnarray}
     \gamma & = \biggl( 1+\frac{\mathcal{E}}{P}\biggr) c_s^{2} .
 \end{eqnarray}\label{gamma}

The coupled differential equations, Eqs.~\eqref{ksi} and \eqref{eta}, include boundary conditions that must be obeyed at the center ($r = 0$):
\begin{equation}
    \eta = -3\gamma \xi 
\end{equation}
and at the surface ($r = R$):
 \begin{equation}
    \eta = \xi \Biggl[ -4 +(1-2M/R)^{-1} \Biggl( -\frac{M}{R} -\frac{\omega^{2} R^{3}}{M}\Biggr)\Biggr] ,
 \end{equation}
with $M$ and $R$ representing the mass and radius of the star. The frequencies are calculated by $ \nu = \omega/{2\pi} = s \: \omega_0/{2\pi}$ (kHz), with $s$ being a dimensionless constant and $\omega_0 \equiv \sqrt{M/R^3}$.

We employ the shooting method to analyze the system, starting the integration with a trial value of $\omega^2$ and initial values that meet the center boundary condition. Integration proceeds toward the surface and the discrete $\omega^2$ values that satisfy boundary conditions correspond to the eigenfrequencies of radial perturbations, described by Sturm-Liouville eigenvalue equations for $\omega$. The solutions yield discrete eigenvalues \ $\omega_n ^{2}$, ordered as
 \begin{equation*}
\omega_0 ^{2} < \omega_1 ^{2} \ldots <\omega_n ^{2}, 
 \end{equation*}
  with $n$ being the number of nodes for a given NS. Stability requires $\omega$ to be real; if any frequency is imaginary, the star becomes unstable.
  To assess stability, it is crucial to determine the fundamental $f$-mode frequency ($n = 0$).
  
\section{Numerical results and discussion}
\label{results}
 For the present work, we study two distinct parameter configurations for each model (NL, NL-$\sigma$ cut, and NL DM), specifically selected to represent the extreme cases of soft and stiff EoS, encompassing the full spectrum of possibilities. All parameter values are presented in Table \ref{tab:nl_models}.

After constructing the EoSs, we proceed to solve the Tolman-Oppenheimer-Volkoff (TOV) equations along with the equations for non-radial and radial oscillations. This approach allows us to determine the stellar characteristics and oscillation patterns for each EoS.

\subsection{EoS and MR Profile}
\label{EoS}

\begin{figure*}[htbp!]
		\begin{minipage}[t]{0.45\textwidth}		 		
  \includegraphics[width=\textwidth]{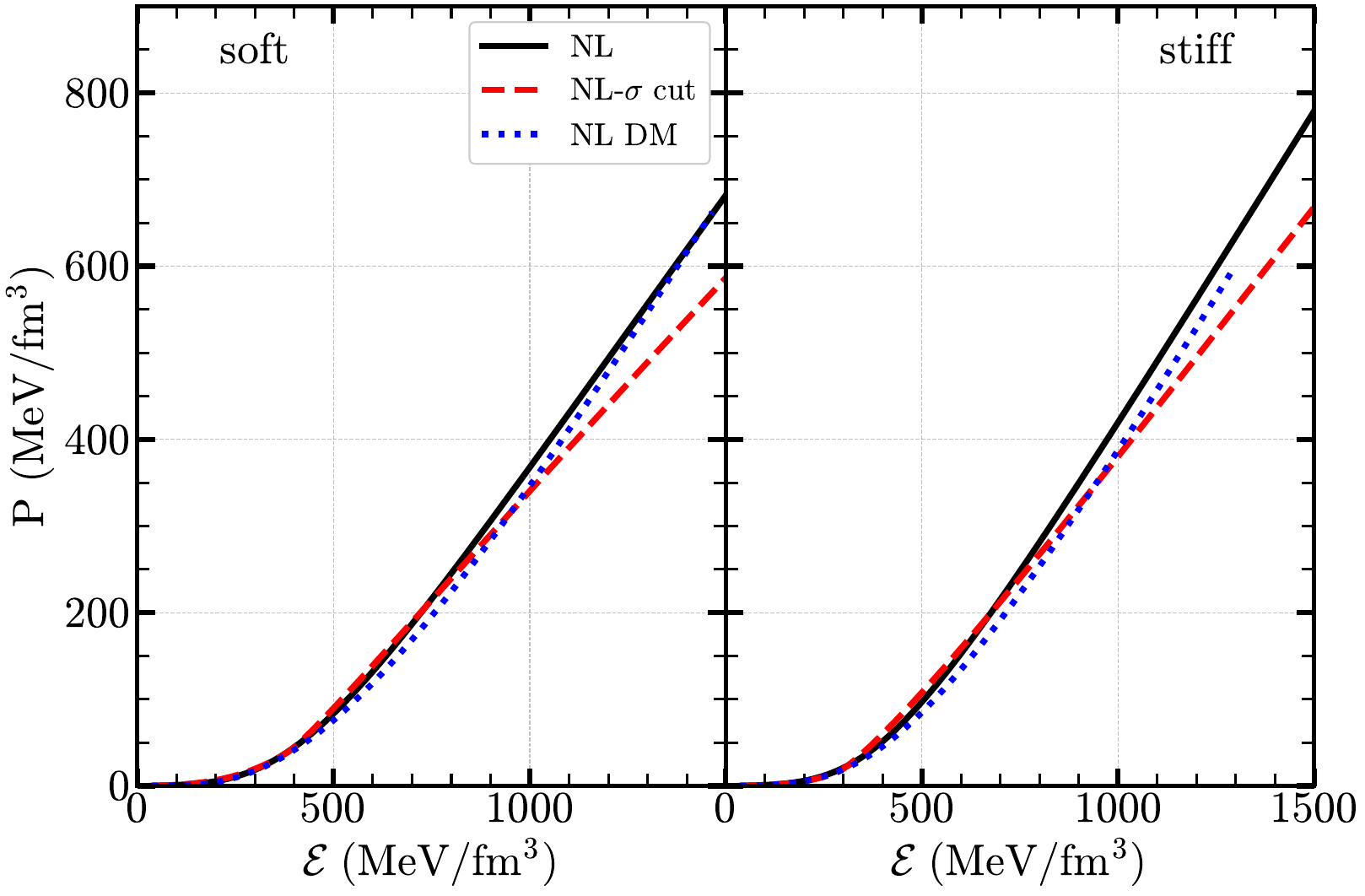}
			 	\end{minipage}
		 		\begin{minipage}[t]{0.50\textwidth}
			 		\includegraphics[width=\textwidth]{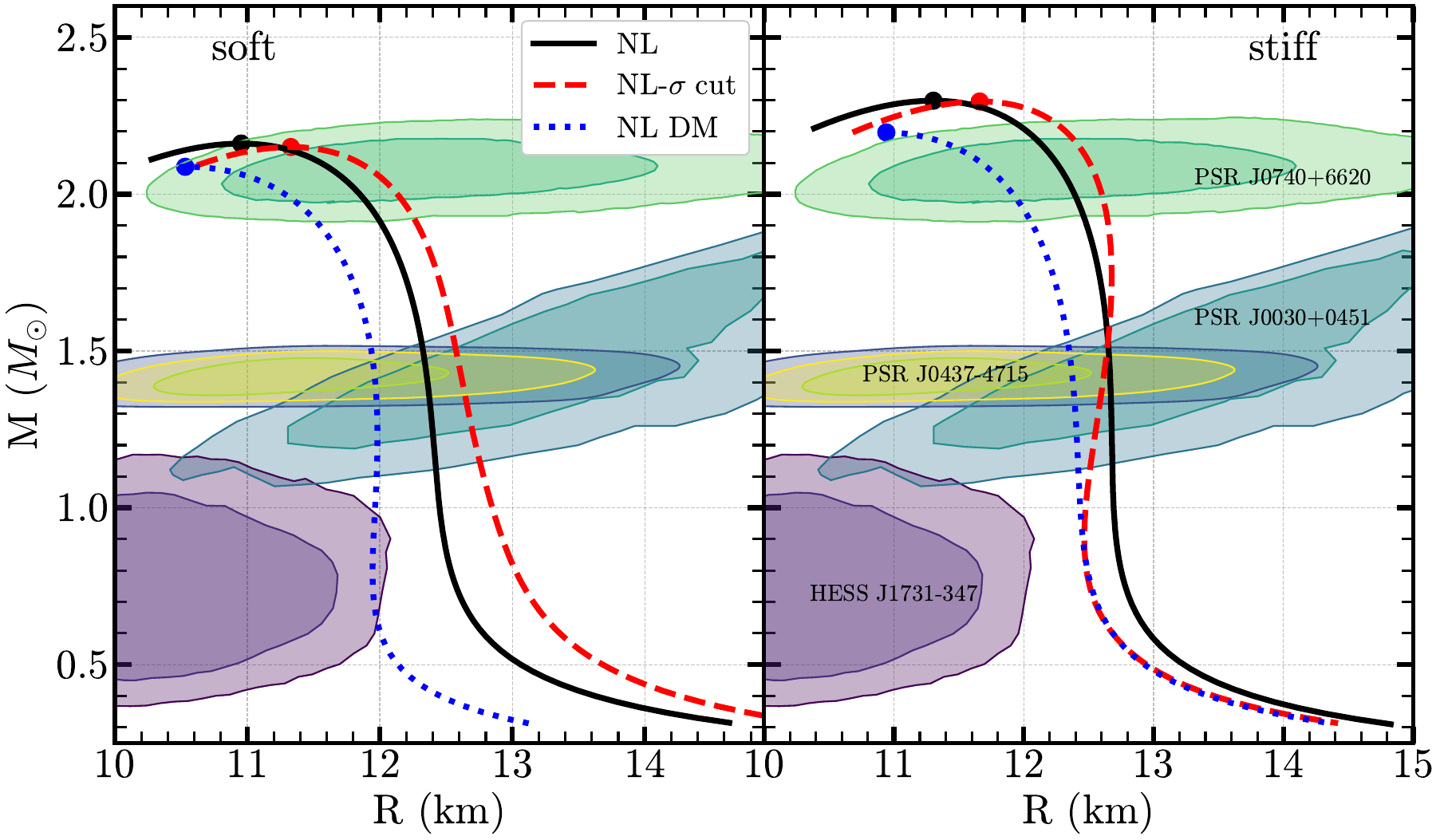}
			 	\end{minipage}
			 			\caption{Left: Energy density and pressure variation for mesonic nonlinear interaction (NL), a modified sigma cut potential (NL-$\sigma$ cut), and dark matter in the NL (NL DM). Right: Mass-Radius (MR) relation for the different EoSs shown. The solid dot represents the last stable point reached in the center of the maximum-mass solution of the TOV equation. The various shaded areas are credibility regions from the mass and radius inferred from the analysis of PSR J0740+6620, PSR J0030+0451, PSR J0437-4715, and HESS J1731-347~\cite{miller2021,2021ApJ...918L..27R, Miller_2019a, Riley_2019, Choudhury:2024xbk, Doroshenko2022}. Each plot's left (right) panel shows the soft (stiff) EoS at different parameter values as discussed in the text. }
		\label{fig:model}	 	
     \end{figure*}
     
Fig.~\ref{fig:model} (left plot) shows the energy density vs. pressure variation for EoS with mesonic nonlinear interaction (NL), a modified sigma cut potential (NL-$\sigma$ cut), and dark matter in the NL (NL DM). The left panel corresponds to the soft EoS, while the right panel corresponds to the stiff ones with the parameter values shown in Table~\ref{tab:nl_models}. From the left panel, we observe that the NL-$\sigma$ cut EoS remains stiff at low to intermediate densities but softens significantly at higher densities. This behavior arises because the $\sigma$-cut modification enhances the stiffness at intermediate densities due to the additional sigma function. Consequently, the requirement for the $\omega$ meson's contribution to support high-mass NSs is reduced. As a result, the coupling constant $g_{\omega}$ is lower in this case, whereas the $\omega^4$ term coupling is larger, leading to an overall softening of the EoS at very high densities. The NL DM EoS remains soft over the whole density range as the addition of DM softens the EoS, but shows increased pressure at very high density. The NL EoS remains the stiffest among the three. From the right panel, the same trend follows as for the left panel, with a stiffer NL EoS compared to the other two. The NL-$\sigma$ cut EoS remains stiffer up to an energy density of $\approx 700 \,\rm{MeV/fm}^3$ which is lower compared to the soft case where it remained stiff until $\approx 650\,\rm{MeV/fm}^3$. 
In our study, we employ a non-unified approach where the low-density crust is modeled using the BPS EoS \cite{Baym:1971pw, Negele:1971vb} and matched to the RMF-based core EoSs at a transition energy density of approximately $ 120\,\rm{MeV/fm}^3$. The matching is carried out ensuring continuity in pressure and chemical potential to preserve thermodynamic consistency \cite{Rather:2020gja, Fortin:2016hny}. Although this approach may introduce minor discontinuities in the speed of sound, we find that it has negligible impact on the global properties and the oscillation mode frequencies reported here, which are predominantly governed by the high-density core.

Solving the TOV equations \cite{PhysRev.55.364, PhysRev.55.374}, we obtain the mass-radius relationship for different EoSs, as shown in Fig.~\ref{fig:model} (right plot). The left (right) panel corresponds to the soft (stiff) EoS at different parameter values. The solid dot on each line represents the last stable point reached in the center of the maximum-mass solution of the TOV equation. The various shaded areas are credibility regions from the mass and radius inferred from the analysis of PSR J0740+6620, PSR J0030+0451, PSR J0437-4715, and HESS J1731-347~\cite{miller2021,2021ApJ...918L..27R, Miller_2019a, Riley_2019, Choudhury:2024xbk, Doroshenko2022}.

In the left panel, the standard NL curve establishes the baseline MR relationship. When the $\sigma$-cut is implemented (NL-$\sigma$ cut), we observe a systematic shift toward larger radii for equivalent masses compared to the standard NL model. This increase in radius indicates decreased compactness, suggesting that the $\sigma$ meson interaction modification leads to reduced pressure in the core, allowing the star to expand to larger radii while supporting similar masses.
The NL DM curve, incorporating dark matter contributions, shows an even more pronounced deviation from the standard NL model. The presence of dark matter further reduces the radius for a given mass, producing more compact stellar configurations. This behavior arises because dark matter adds to the total energy density while contributing relatively little pressure, effectively softening the EoS. As a result, the pressure support against gravity is reduced, leading to increased compactness without an accompanying increase in pressure support, such as that provided by degeneracy pressure in ordinary matter. 
In the right panel, we focus exclusively on the stiff EoS regime. Here, the standard NL curve again serves as our reference point. When implementing the $\sigma$-cut mechanism (NL-$\sigma$ cut), we observe a consistent shift toward smaller radii up to a mass of around $1.5\,M_{\odot}$, which then increases until the maximum mass.

The presence of dark matter, NL DM, produces a systematic shift toward smaller radii for equivalent masses when compared to both the standard NL and NL-$\sigma$ cut models. Upto a mass of $\approx 1.0\,M_{\odot}$, the MR curve for the NL DM EoS overlaps with the NL-$\sigma$ cut one, beyond which the NL DM curve decreases to a lower radii, leading to a smaller maximum mass configuration as compared to the NL-$\sigma$ cut, because of the softening of the EoS by DM. 
More details regarding this behavior can be seen in Ref.~\cite{Thakur:2024btu}

These systematic modifications to the mass-radius relationship demonstrate how sensitive NS structure is to variations in the underlying nuclear physics assumptions and compositional models, even when remaining within the broader stiff EoS category.
The observational constraints appear to favor the intermediate models rather than the extremes, suggesting that modifications to the standard NL model may indeed be necessary to accurately describe the NS structure. Particularly, the constraint from PSR J0740+6620, with its relatively high mass, provides an important test for these models' predictions in the high-mass regime. The stellar properties such as the maximum mass ($M_{\rm max}$), radius at the maximum mass ($R_{\rm max}$), at $2.0\,M_{\odot}$ ($R_{2.0}$), and at $1.4\,M_{\odot}$ ($R_{1.4}$) along with the dimensionless tidal deformability ($\Lambda_{1.4}$) are shown in Table~\ref{tab:prop}. All the EoSs, soft as well as stiff, satisfy the necessary astrophysical constraints from mass, radius, and tidal deformability \cite{miller2021,2021ApJ...918L..27R, Miller_2019a, Riley_2019, Choudhury:2024xbk, Doroshenko2022, LIGOScientific:2018cki}.

\begin{table*}[htb!]
\centering
\caption{Stellar properties for different compositions of EoS: maximum mass ($M_{\text{max}}$), radius at maximum mass ($R_{\text{max}}$), at 2.0\,$M_{\odot}$ ($R_{2.0}$), and at 1.4\,$M_{\odot}$ ($R_{1.4}$). Dimensionless tidal deformability at 1.4\,$M_{\odot}$ ($\Lambda_{1.4}$), non-radial $f$-mode frequency $f_{non-radial}$, radial $f$-mode frequency $f_{radial}$, and damping time $\tau_{f}$ (for non-radial) at 1.4\,$M_{\odot}$.}
\begin{tabular}{ccccccccc}
\hline
Composition & $M_{\text{max}}$ ($M_{\odot}$) & $R_{\text{max}}$ (km) & $R_{2.0}$ (km) & $R_{1.4}$ (km) & $\Lambda_{1.4}$ & $f_{non-radial}$ (kHz) & $f_{radial}$ (kHz) & $\tau_{f}$ (sec) \\
\hline
NL soft & 2.16 & 10.95 & 11.84 & 12.36 & 408 & 1.81 & 1.62 & 0.217 \\
\hline
NL stiff & 2.29 & 11.30 & 12.39 & 12.66 & 486 & 1.74 & 1.41 & 0.235 \\
\hline
NL-$\sigma$ cut soft & 2.15 & 11.33 & 12.13 & 12.61 & 381 & 1.83 & 1.65 & 0.212 \\
\hline
NL-$\sigma$ cut stiff & 2.30 & 11.66 & 12.60 & 12.62 & 508 & 1.73 & 1.44 & 0.240 \\
\hline
NL DM soft & 2.09 & 10.53 & 11.29 & 11.97 & 369 & 1.85 & 1.69 & 0.209 \\
\hline
NL DM stiff & 2.19 & 10.94 & 11.91 & 12.38 & 420 & 1.80 & 1.48 & 0.219 \\
\hline
\end{tabular}
\label{tab:prop}
\end{table*}

\subsection{Non-radial profiles}

\begin{figure}[h]		 		
  \includegraphics[width=0.47\textwidth]{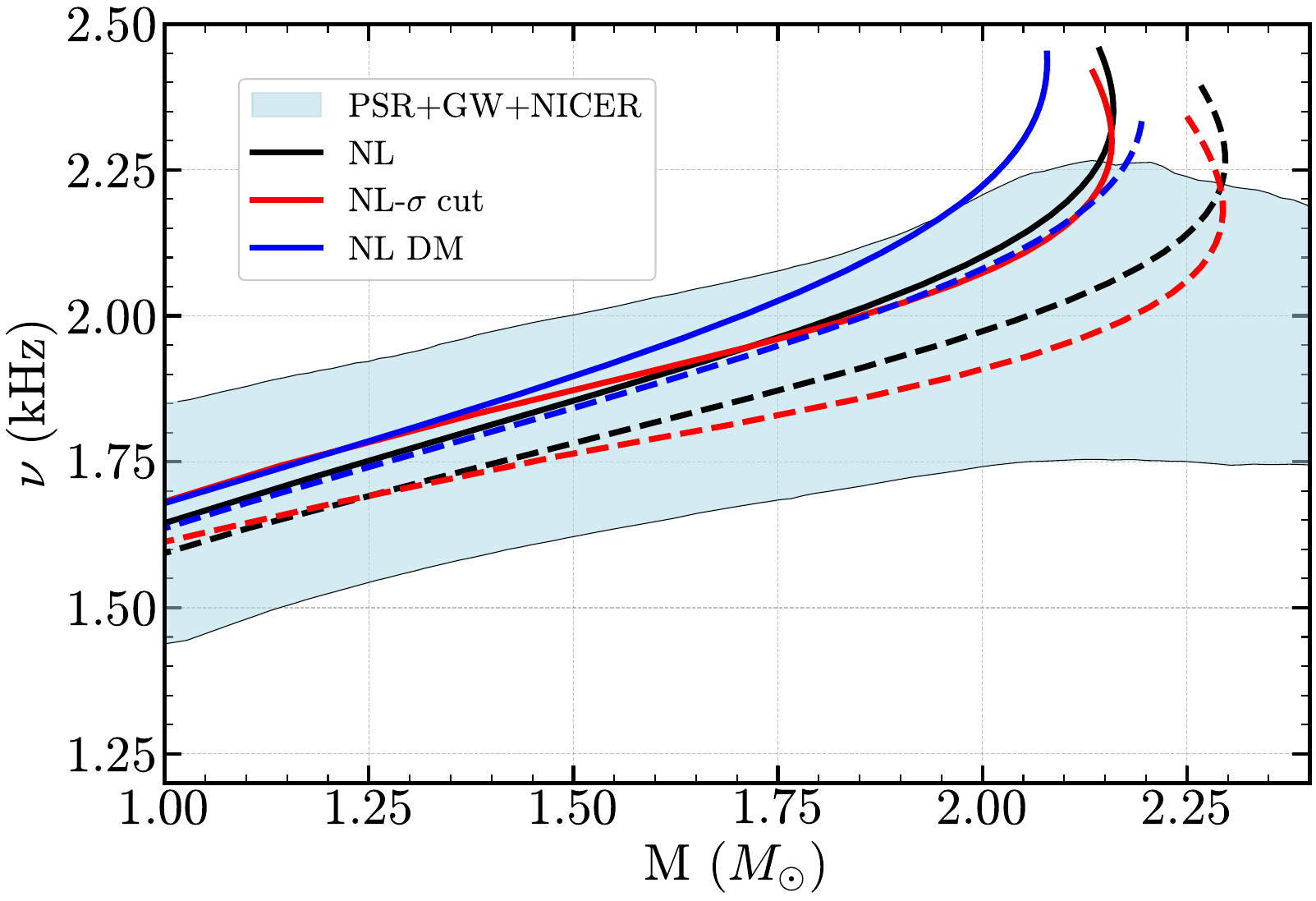}
			 			\caption{ Non-radial $f$-mode frequency as a function of stellar mass for NL, NL-$\sigma$ cut, and NL DM EoS. The solid (dashed) lines correspond to the soft (stiff) EoSs. All the calculations are performed within a full general relativistic (GR) framework. The blue shaded region encapsulates the 90\% credible interval on $f$-mode frequency from PSR+GW+NICER \cite{Mohanty:2024usv}. }
		\label{fignR}	 	
     \end{figure}
     
Fig.~\ref{fignR} represents the $f$-mode frequencies as a function of neutron star mass for NL, NL-$\sigma$ cut, and NL DM EoS. Solid lines indicate soft EoSs, while dashed lines denote stiff EoSs. Among the soft EoSs, the NL DM model exhibits the highest frequency of 1.85 kHz at the canonical mass of 1.4\,$M_{\odot}$, exceeding the frequencies of NL and NL-$\sigma$ cut, as shown in Table~\ref{tab:prop}. This increase in frequency may be attributed to the presence of dark matter in the NL DM EoS, which leads to a softening of the EoS. In contrast, stiff EoSs produce lower $f$-mode frequencies at 1.4\,$M_{\odot}$ compared to soft EoSs, as indicated in Table \ref{tab:prop}. At $M_{\text{max}}$, the NL DM soft model attains the highest frequency of 2.43~kHz, while the NL-$\sigma$ cut model exhibits the lowest frequency of 2.17~kHz, as clearly shown in Fig.~\ref{fignR}. Our results are in agreement with work by \citet{Shirke:2024ymc} for the NL DM case, where the authors studied the same DM model using a full GR framework for the $f$-mode, and also with \citet{Kalita:2024kbv}, where the $\sigma$-cut model was investigated for the $f$-mode using the Cowling approximation. The shaded region labeled PSR+GW+NICER represents the 90\% credible interval for $f$-mode frequencies derived using a nonparametric EoS framework conditioned on astrophysical data from pulsar mass measurements, gravitational wave (GW170817) tidal deformabilities, and NICER radius observations \cite{Mohanty:2024usv}.
Our EoS models perfectly lie within this credible band at both $1.4$ and $2.0\,M_\odot$. The soft and stiff EoS of NL and NL-$\sigma$ cut satisfy the constraints throughout. The NL DM soft EoS falls well within the band at $1.4\,M_\odot$, while at $2.0\,M_\odot$ it marginally exceeds the upper boundary. 
This agreement confirms that our EoSs are consistent with current multimessenger constraints and support realistic dynamical properties of neutron stars.
\begin{figure}[t]		 		
  \includegraphics[width=0.47\textwidth]{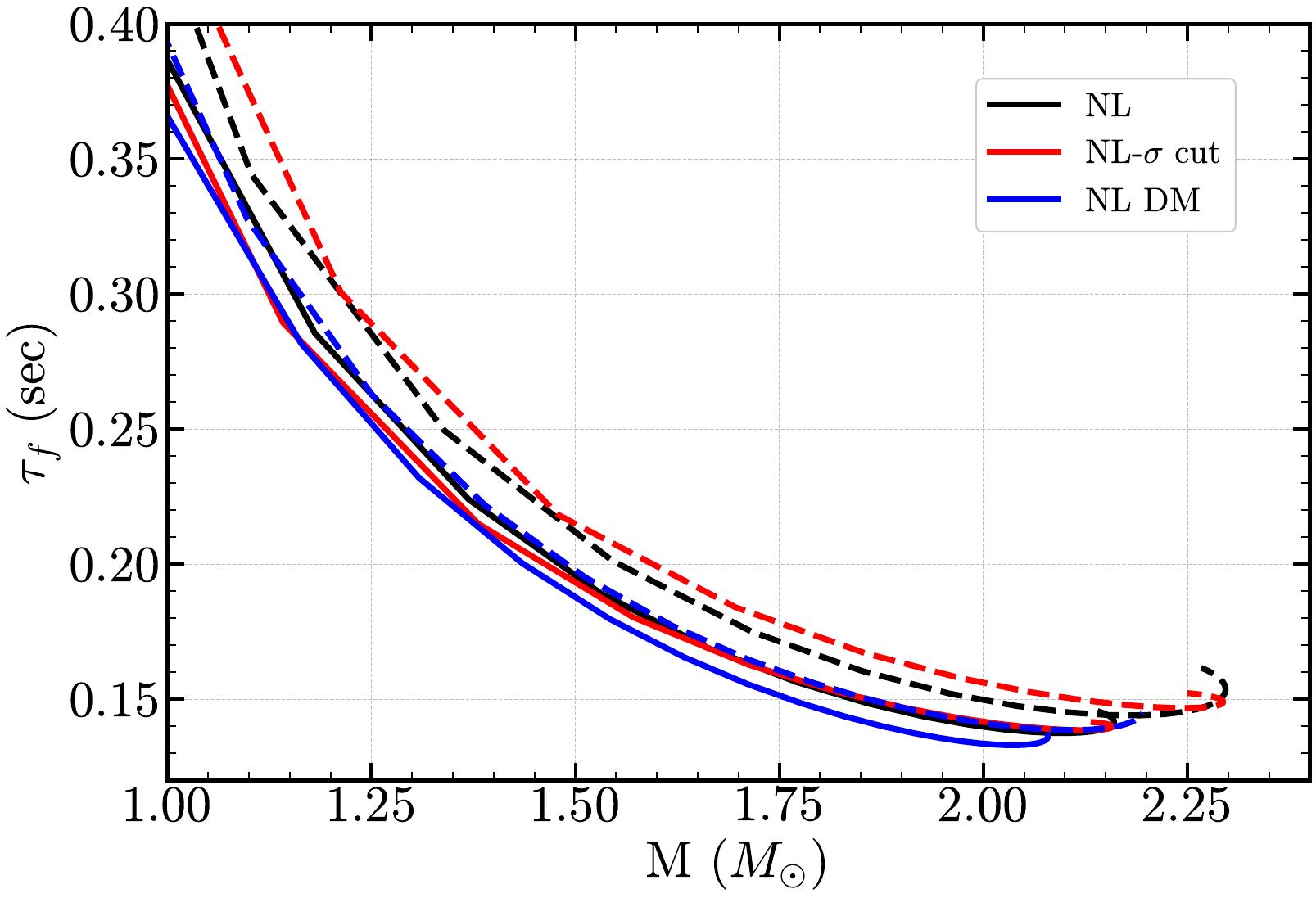}
			 			\caption{The $f$-mode frequency damping time $\tau_f$ (in sec.) as a function of stellar mass for NL, NL-$\sigma$ cut, and NL DM EoS. The solid (dashed) lines correspond to the soft (stiff) EoSs.  }
		\label{figdamp}	 	
     \end{figure}
     
These patterns in $f$-mode frequencies can be better understood by considering the role of compactness, which is influenced by whether the EoS is soft or stiff. Soft EoSs yield more compact NSs with smaller radii, which affects the behavior of $f$-mode oscillations. Since they are characterized by fluid perturbations peaking near the stellar surface and metric perturbations peaking at the center \cite{Kunjipurayil:2022zah}, the compactness of a star plays a crucial role. We have also computed the damping time of $f$-modes, with its dependence on the NS mass depicted in Fig.~\ref{figdamp}. For a typical NS, the $f$-mode frequency falls within the range of 1–3 kHz, while the corresponding damping time, $\tau_f$, is typically a few tenths of a second \cite{Kokkotas:1999bd,Wen:2019ouw,Pradhan:2020amo}. As the NS mass increases, the damping time steadily decreases. The variations in the curves reflect the properties of the softest and stiffest EoSs, as summarized in Table~\ref{tab:prop}. Soft EoSs exhibit the shortest damping times, ranging from 0.209 to 0.217 seconds, while the stiff EoSs have the damping time in the range 0.219-0.240 seconds. For NSs with stiff EoS, the larger radius leads to a lower mean density, weakening the restoring force for fluid oscillations and resulting in lower $f$-mode frequencies. Additionally, the lower compactness reduces metric perturbations and gravitational wave damping, leading to longer damping times. In contrast, for NSs with a soft EoS, the smaller radius leads to a higher mean density, strengthening the restoring force and increasing the $f$-mode frequency. The higher compactness enhances metric perturbations, leading to shorter damping times, which in turn enables significant emission of gravitational radiation \cite{Lindblom:1983ps,PhysRevD.84.044017}.

\begin{figure}[h]		 		
  \includegraphics[width=0.47\textwidth]{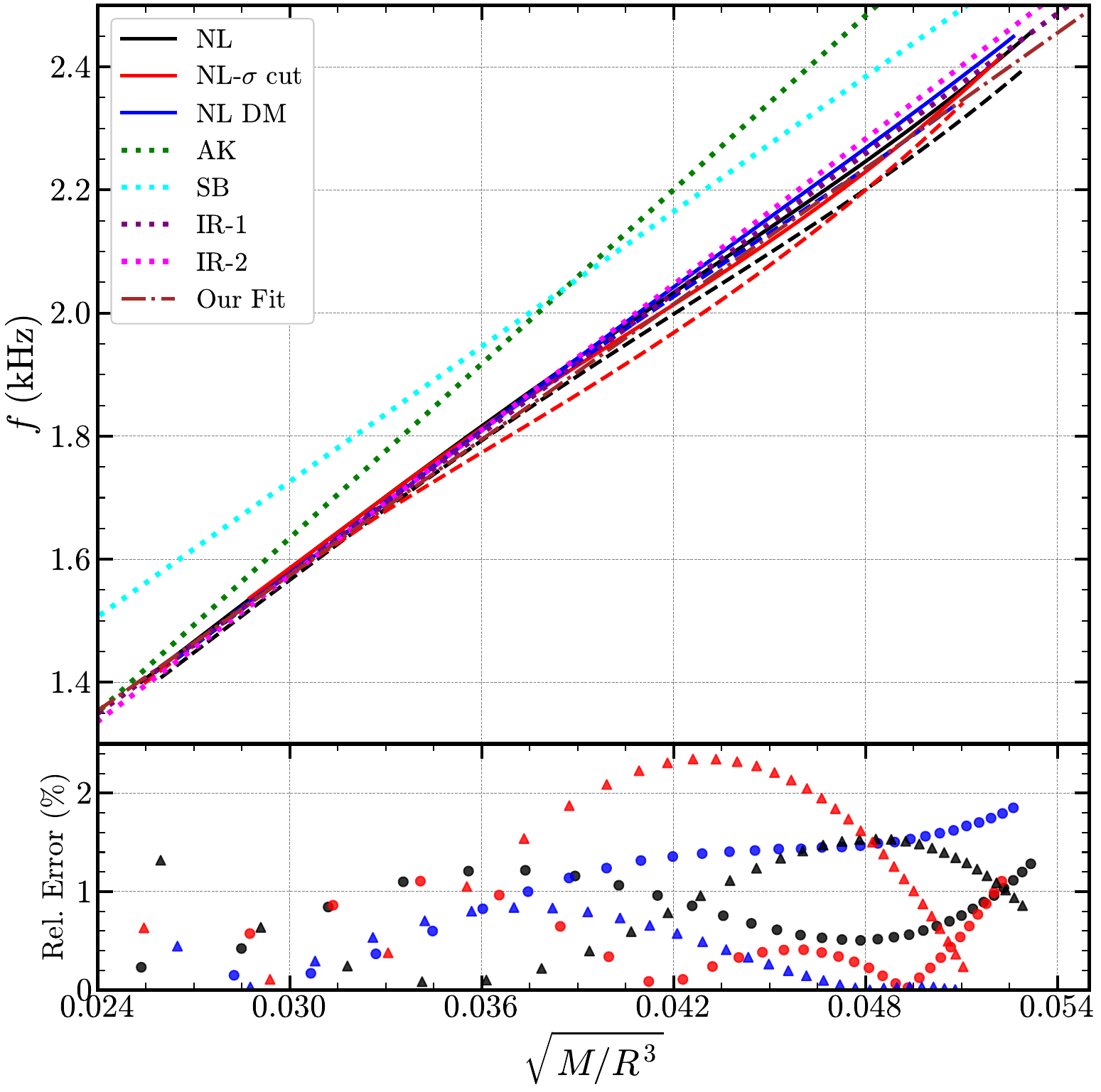}
			 			\caption{The $f$-mode frequency as a function of the mean stellar density $\sqrt{M/R^3}$ for NL, NL-$\sigma$ cut, and NL DM EoS. The solid (dashed) lines correspond to the soft (stiff) EoSs. Several different fits from previous studies are shown by dotted lines, while the brown dash-dotted line represents our fit. The lower panel shows the relative error (\%) between the data and our best-fit line, where solid markers denote soft EoSs and triangular markers denote stiff EoSs.}  
		\label{figden_f}	 	
     \end{figure}

\begin{figure}[h]		 		
  \includegraphics[width=0.47\textwidth]{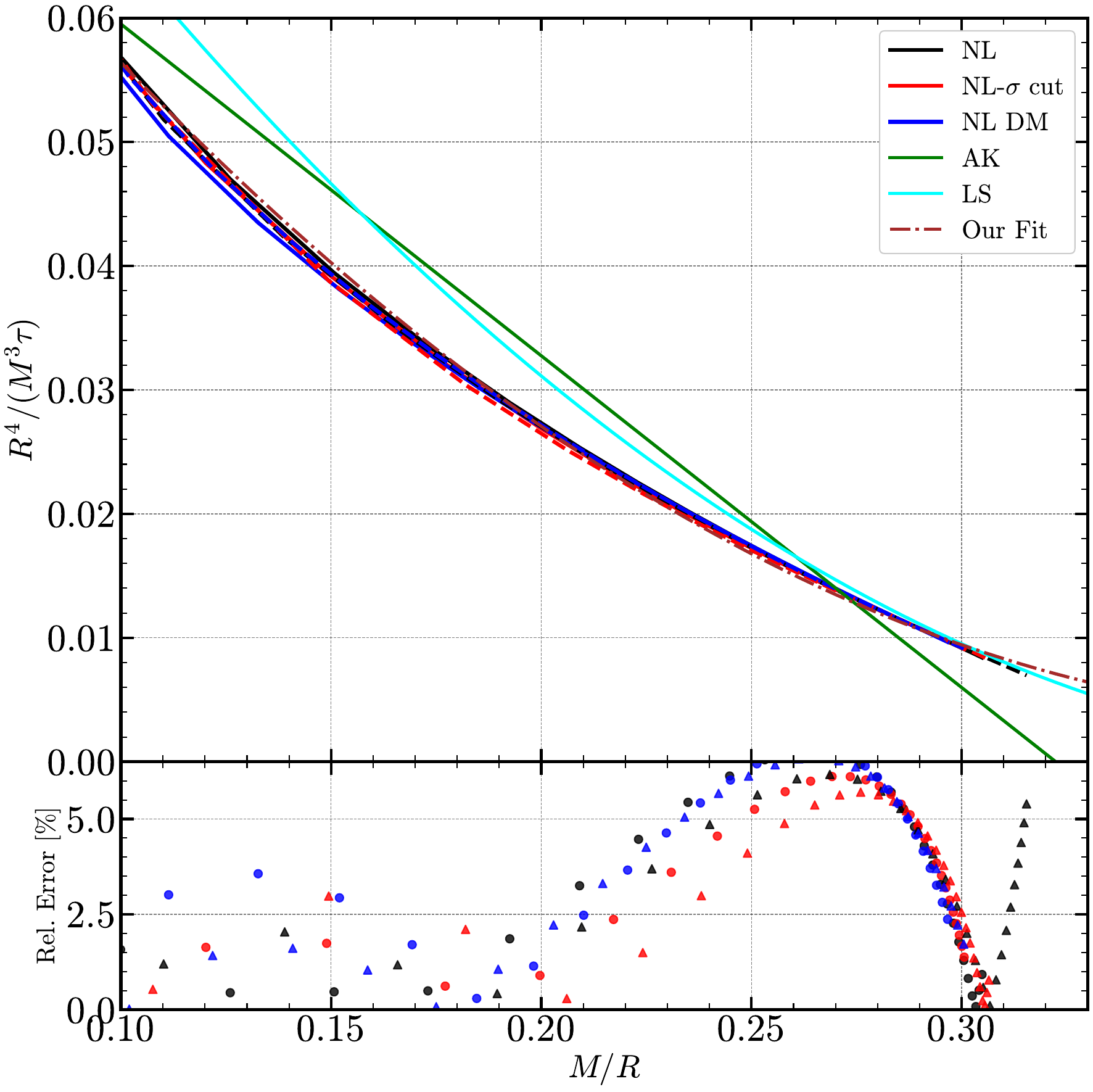}
			 			\caption{Normalized damping time $R^4/(M^3 \tau)$ of the $f$-mode as a function of the stellar compactness $C=M/R$. The solid (dashed) lines correspond to the soft (stiff) EoSs. The green and cyan lines are fitted from previous studies, AK \cite{Andersson:1997rn} and LS \cite{Lioutas:2017xtn}, respectively, while the brown dash-dotted line represents our fit. The lower panel shows the relative error (\%) between the data and our best-fit line, where solid markers denote soft EoSs and triangular markers denote stiff EoSs.} 
		\label{figC_f}	 	
     \end{figure}

Next, we will focus on quasi-universal relation, which are nearly independent of the EoS and play a crucial role. This is because extracting NS properties directly from gravitational wave frequency observations is challenging due to the uncertainty in the NS matter EoS. So far, several quasi-universal relation have been established, including the fundamental $f$-mode frequency as a function of the NSs average density, as a function of compactness, and the spacetime $w$ mode frequency and damping rate in relation to compactness \cite{Benhar:1998au,Sotani:2020bey,Lau:2009bu}. If the characteristics of gravitational waves are observationally determined, these quasi-universal relations can aid in constraining key macroscopic properties of NSs—such as mass, radius, and compactness—independently of the underlying microphysics. These properties, in turn, may be used, together with additional observational data, to constrain the EoS of dense matter indirectly.

A widely recognized relation, proposed by \citet{Andersson:1997rn}, describes the connection between the $f$-mode frequency and the star's mean density, expressed as:
\begin{equation}
\frac{f}{\text{kHz}} \approx 0.22 + 32.16 \sqrt{\frac{M}{R^3}}.
\label{eq:f_mode_relation}
\end{equation}
This relation is shown in a green dotted line with legend AK in Fig.~\ref{figden_f}. The figure illustrates the relation between the $f$-mode frequency $f$ (in kHz) and the square root of the mean stellar density $\sqrt{M/R^3}$ for different EoSs: NL (black), NL-$\sigma$ cut (red), and NL DM (blue). Solid lines correspond to softer EoSs, while dashed lines represent stiffer variants.

\citet{Shirke:2024ymc} proposed a similar empirical fit for DM-admixed NSs, represented by the cyan dotted line (SB) in the figure. 

\begin{equation}  
\frac{f}{\text{kHz}} \approx 0.630 + 33.54 \sqrt{\frac{M}{R^3}}.  
\label{eq:f_mode_bikram}  
\end{equation}  
However, the relation between $f$ and $M/R^3$ exhibits a degree of model dependence, as also noted in Refs.~\cite{Pradhan:2020amo,Pradhan:2022vdf}. In our analysis, we perform an independent fit for the NL, NL-$\sigma$ cut, and NL DM models. The resulting fit is given by:

\begin{equation}
\frac{f}{\text{kHz}} \approx 0.473 + 36.706 \sqrt{\frac{M}{R^3}}.
\label{eq:f_mode_relation_our}
\end{equation}
which can be seen in Fig.~\ref{figden_f} by the brown dash-dotted line (Our Fit). The lower panel displays the relative error (\%) between the calculated $f$-mode frequencies from the data and the best-fit line proposed in this work. Each marker corresponds to a NS configuration based on a given EoS. Solid markers represent soft EoS, while triangular markers denote stiff EoSs. The plot shows that the relative error remains below approximately 2.5\% throughout the range of mean stellar densities considered. This indicates a strong agreement between the fit and the numerical data across all three EoS models analyzed. The clustering of most points below the 2\% threshold supports the quasi-universal character of the $f$--$\sqrt{M/R^3}$ relation, even in the presence of dark matter and $\sigma$-cut potential modifications to the EoS.

While all fits exhibit a near-linear trend, clear deviations are evident. Notably, both the AK and SB fits lie systematically above the $f$-mode frequencies obtained from our models across most of the $\sqrt{M/R^3}$ range. This suggests that the previous empirical relations, while broadly capturing the trend, do not fully account for the EoS-dependent features introduced by additional physics such as dark matter or modified sigma field.
In our previous work \cite{Rather:2024nry}, we established two quasi-universal relations between $f$-mode frequency and mean stellar density for different classes of exotic matter. We derived fit parameters IR-1 (purple dotted line; $a=0.44, b=37.90$) for hyperons and $\Delta$ baryons without phase transitions, and IR-2 (magenta dotted line; $a=0.39, b=39.44$) for EoSs incorporating phase transitions to quark matter. Remarkably, these previously obtained fits perfectly describe our current data that includes dark matter, demonstrating the robust universality of these relations across diverse compositions. IR-1 fits more perfectly as compared to IR-2, especially at the high-density region. The reason could be due to the phase transition to the quark matter for IR-2. While other fits shown in the plot capture specific cases, IR-1 and IR-2 demonstrate superior predictive power by accurately characterizing our new dark matter results without requiring additional parameters. This strong agreement between our earlier parameterizations and present calculations suggests that the underlying relations are remarkably robust across a variety of microphysical scenarios, including those with hyperons, $\Delta$ baryons, phase transitions, and dark matter components. While not claiming true universality, e.g, in the strict I-Love-Q sense \cite{Yagi:2013awa, Yagi:2013bca}, our results indicate that these relations retain their form even when applied to significantly modified EoSs within the RMF framework.

Fig.~\ref{figC_f} presents the normalized damping time of the $f$-mode, given by $R^4/(M^3 \tau)$, as a function of the compactness $M/R$, allowing direct comparison between the models used in this work and existing empirical fits. As compactness increases, the normalized damping time decreases monotonically, indicating stronger gravitational wave damping in more compact stars.

\citet{Andersson:1997rn} initially proposed a quasi-universal relation for the $f$-mode damping time $\tau$. This relation was later extended by \citet{Lioutas:2017xtn} to include higher-order terms, resulting in the following expression:
\begin{equation}
\frac{R^4}{M^3 \tau} \approx 0.112 - 0.53 \frac{M}{R} + 0.628 \left(\frac{M}{R} \right)^2.
\end{equation}
This extended fit, shown as the cyan curve in Fig.~\ref{figC_f}, attempts to capture the nonlinear dependence of the normalized damping time on stellar compactness. 

In our case, this quadratic fit for all three models is represented by the brown dash-dotted line in Fig.~\ref{figC_f}, which reads as:
\begin{equation}
\frac{R^4}{M^3 \tau} \approx 0.097 - 0.469 \frac{M}{R} + 0.586 \left(\frac{M}{R} \right)^2.
\end{equation}

The lower panel shows the relative error (\%) between the numerical data and the empirical best-fit line derived in this work. Solid markers represent soft EoSs, while triangular markers correspond to stiff EoSs. Across most of the compactness range, the relative error remains below 5\%, indicating a high degree of agreement between the data and the fit. Notably, the error slightly increases for higher compactness values, particularly beyond $M/R \sim 0.25$, but remains within acceptable limits. This suggests that the proposed relation captures the overall behavior of the dataset with reasonable accuracy across a wide range of NS configurations.

It is evident that the AK and LS fits deviate noticeably from the data used in this work. This indicates that existing quasi-universal relations, while robust across a broad class of standard hadronic EoSs, may not fully capture the variations introduced by specific microphysical modifications such as the $\sigma$-cut potential and dark matter admixture. The new fit thus provides a more accurate description of the $f$-mode damping times across a wide range of stellar compactness, enhancing the reliability of gravitational wave-based neutron star characterization.


\subsection{$p_1$ mode}

\begin{figure}[htbp]
  \centering
    \includegraphics[width=\linewidth]{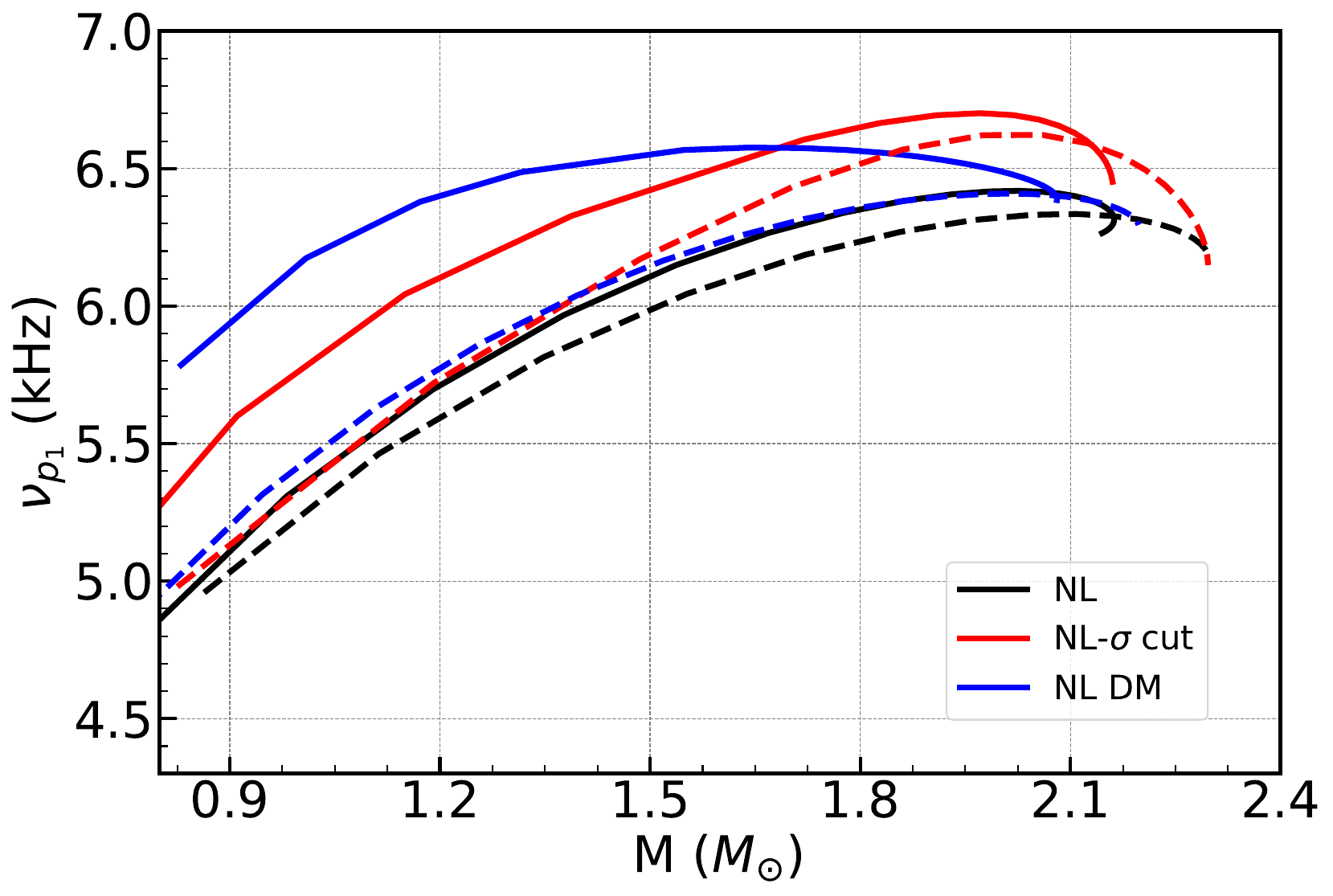}\\
    \includegraphics[width=\linewidth]{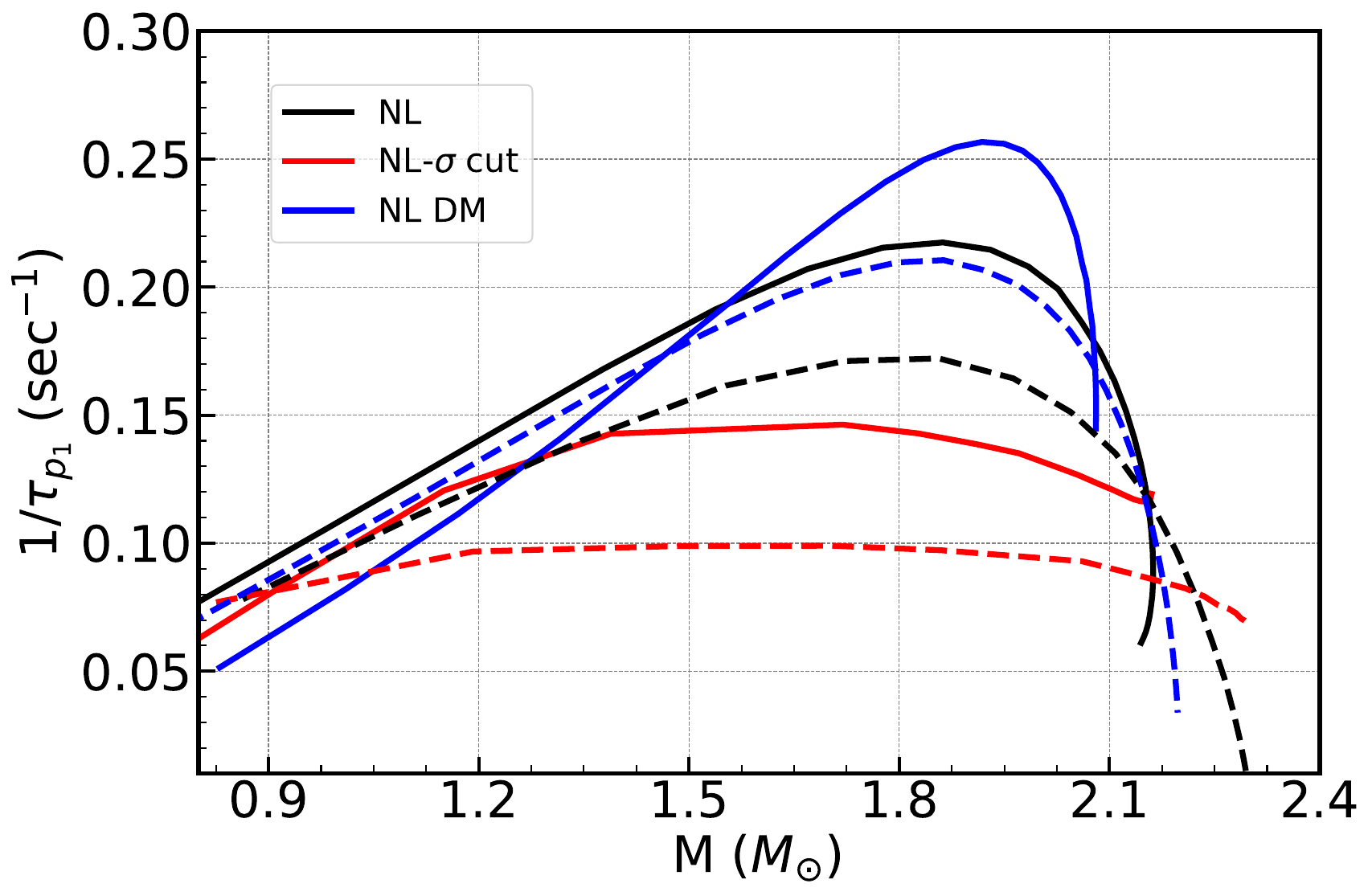}
  \caption{Top: Non-radial $p_1$-mode frequencies as a function of stellar mass for NL, NL-$\sigma$ cut, and NL DM EoS. The solid (dashed) lines correspond to the soft (stiff) EoSs. Bottom: Variation of the imaginary part $\omega$ of $p_1$-mode oscillations (inverse damping time) versus NS mass.}
  \label{fig:damping_time_all}
\end{figure}

To extend our current study of non-radial oscillations, we calculate the non-radial $p_1$-mode for $\ell = 2$. The $p_1$-mode frequencies and their damping times offer important information about the internal structure and oscillation characteristics of NSs across various EoS models \cite{Kunjipurayil:2022zah}. The top panel of Fig.~\ref{fig:damping_time_all} presents the variation of $p_1$-mode frequencies as a function of stellar mass for three different models: NL, NL-$\sigma$ cut, and NL DM. The solid lines correspond to the soft EoSs, while the dashed lines represent the stiff EoSs. The $p_1$-mode frequency spans 
from $\approx$ 5\,kHz to 6.5\,kHz. Across all models, the frequency increases with mass, reaching a peak before slightly decreasing at higher masses. 
At $1.4\,M_{\odot}$, the NL DM soft model shows the highest frequency of 6.52~kHz, followed by the NL-$\sigma$ soft, while the NL stiff model yields the lowest frequency at 5.873~kHz, as shown in Table~\ref{tab:p1}. At $M_{\text{max}}$, the soft NL-$\sigma$ cut model exhibits the highest frequency of 6.452\,kHz, followed closely by the NL DM soft model with a frequency of 6.390\,kHz. To provide a more quantitative comparison, Table~\ref{tab:p1} also lists the corresponding damping times. These follow an inverse trend, with the NL-$\sigma$ stiff model exhibiting the longest damping time of 10.176 seconds and the NL soft model showing the shortest at 5.851 seconds. The bottom panel of Fig.~\ref{fig:damping_time_all} illustrates the variation of the imaginary part of the oscillation frequency, which corresponds to the inverse damping time, $1/\tau$. We found that the damping time of the $p_1$-mode, $\tau_{p_1}$, is significantly longer than that of the $f$-mode, consistent with the findings reported in Ref.~\cite{Kunjipurayil:2022zah, Thakur:2024ijp}. The damping rate increases with mass, peaking around $\sim 2 M_{\odot}$ before declining at higher masses. This trend can be understood from the modifications to the EoS. The inclusion of DM softens the EoS, resulting in more compact stars that oscillate at higher frequencies but lose energy more efficiently, leading to shorter damping times. In contrast, the $\sigma$-cut potential stiffens the EoS by enhancing repulsive interactions, resulting in lower compactness stars. These stars oscillate at lower frequencies but dissipate energy more slowly, leading to extended damping times. The extended damping time in this model is a direct consequence of the enhanced rigidity, which slows down the rate of energy loss. 

The distinct frequency peaks and damping-time behavior for each model underscore how detailed measurements of neutron star quasi-normal modes could serve as sensitive probes of the underlying microphysics, including potential dark matter components or modified nuclear interactions.

\begin{table}[htbp]
  \centering
  \caption{Non-radial frequency $\nu_{p_{1}}$, and damping time $\tau$ at 1.4\,$M_{\odot}$.}
  \begin{tabular}{lcc}
    \hline
    Model              & $\nu_{p_{1}}$ (kHz) & $\tau$ (sec) \\
    \hline
    NL soft      & 5.99          & 5.851              \\
    NL stiff     & 5.87          & 6.894              \\
    NL-$\sigma$ cut soft   & 6.34          & 6.999              \\
    NL-$\sigma$ cut stiff & 6.04         & 10.176             \\
    NL DM soft   & 6.52         & 6.421              \\
    NL DM stiff & 6.04         & 6.111              \\
    \hline
  \end{tabular}
  \label{tab:p1}
\end{table}

\subsection{Radial profiles}
\begin{figure*}[t]
		\begin{minipage}[t]{0.49\textwidth}		 		
  \includegraphics[width=\textwidth]{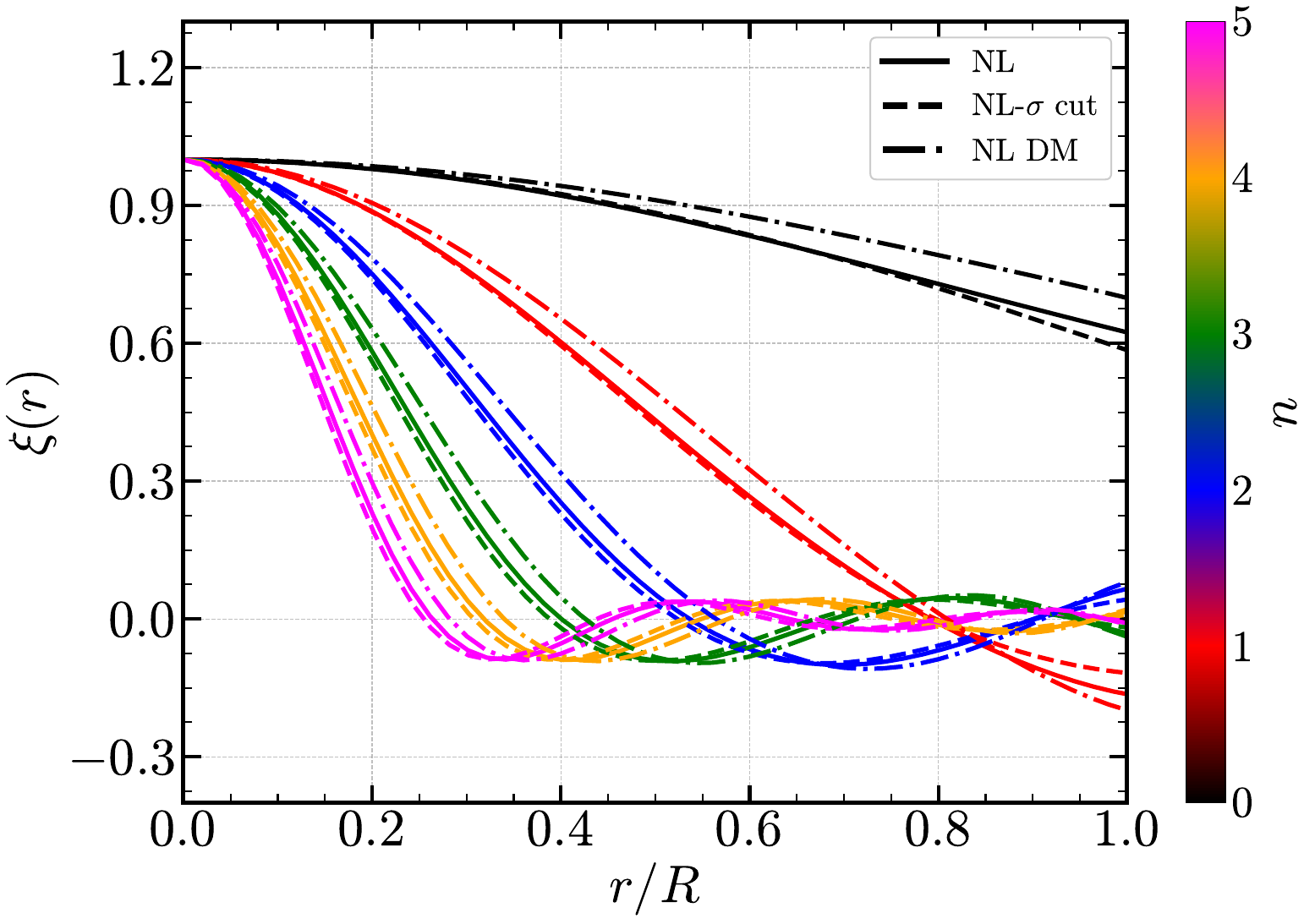}
			 	\end{minipage}
		 		\begin{minipage}[t]{0.49\textwidth}
			 		\includegraphics[width=\textwidth]{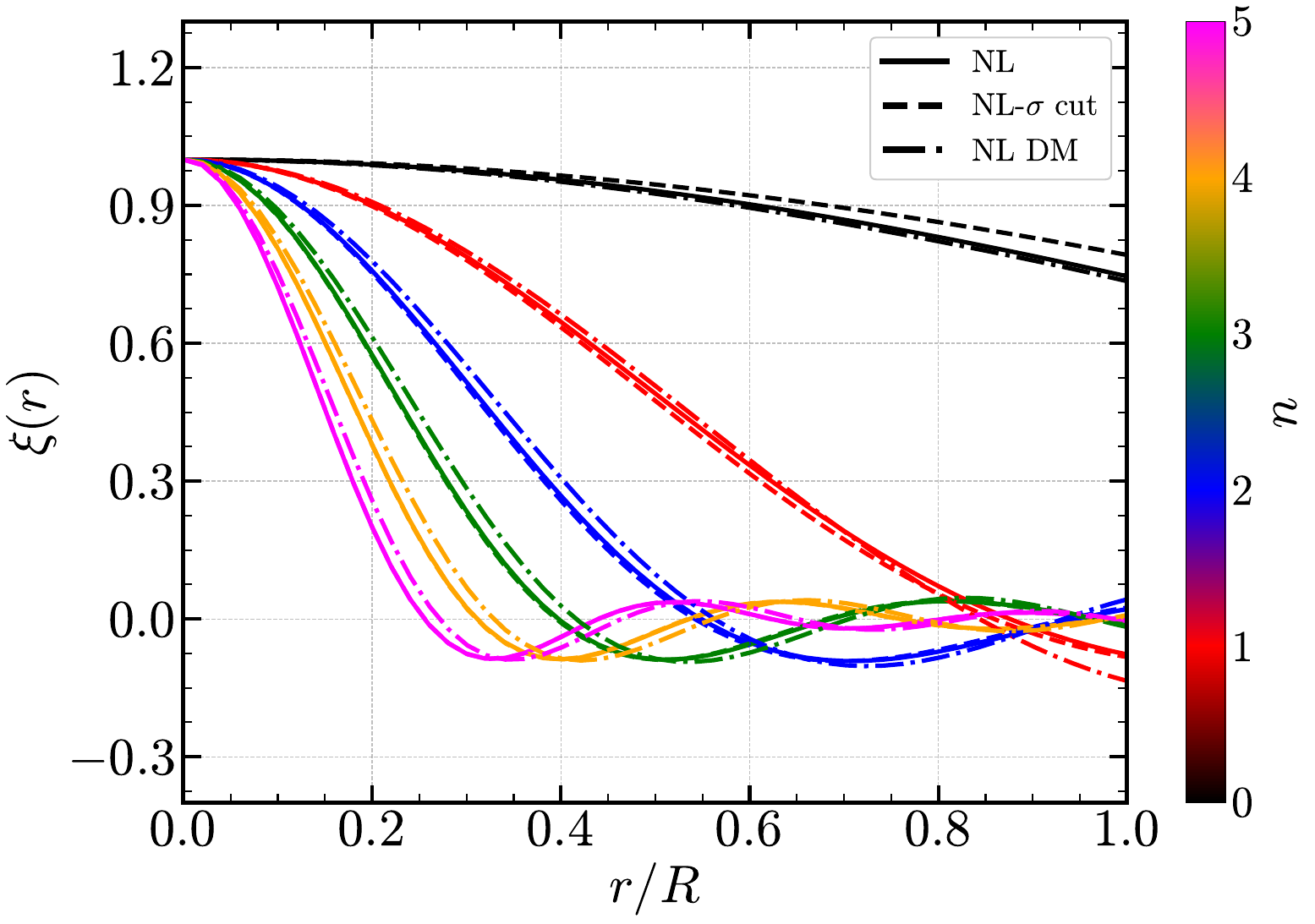}
			 	\end{minipage}
			 			\caption{Left: The radial displacement perturbation ${\xi(r)}$ = ${\Delta r/r}$ as a function of dimensionless radius distance ${r/R}$. The colorbar represents the modes: ${f}$-mode ($n = 0$) and lower-order ${p}$-modes ($n = 1-5$) for soft NL, NL-$\sigma$ cut, and NL DM EoS. Right: Same as left plot, but for stiff EoSs. All the modes are calculated at 1.4\,${M_{\odot}}$. }
		\label{fig:ksi}	 	
     \end{figure*}

\begin{figure*}[t]
		\begin{minipage}[t]{0.49\textwidth}		 		
  \includegraphics[width=\textwidth]{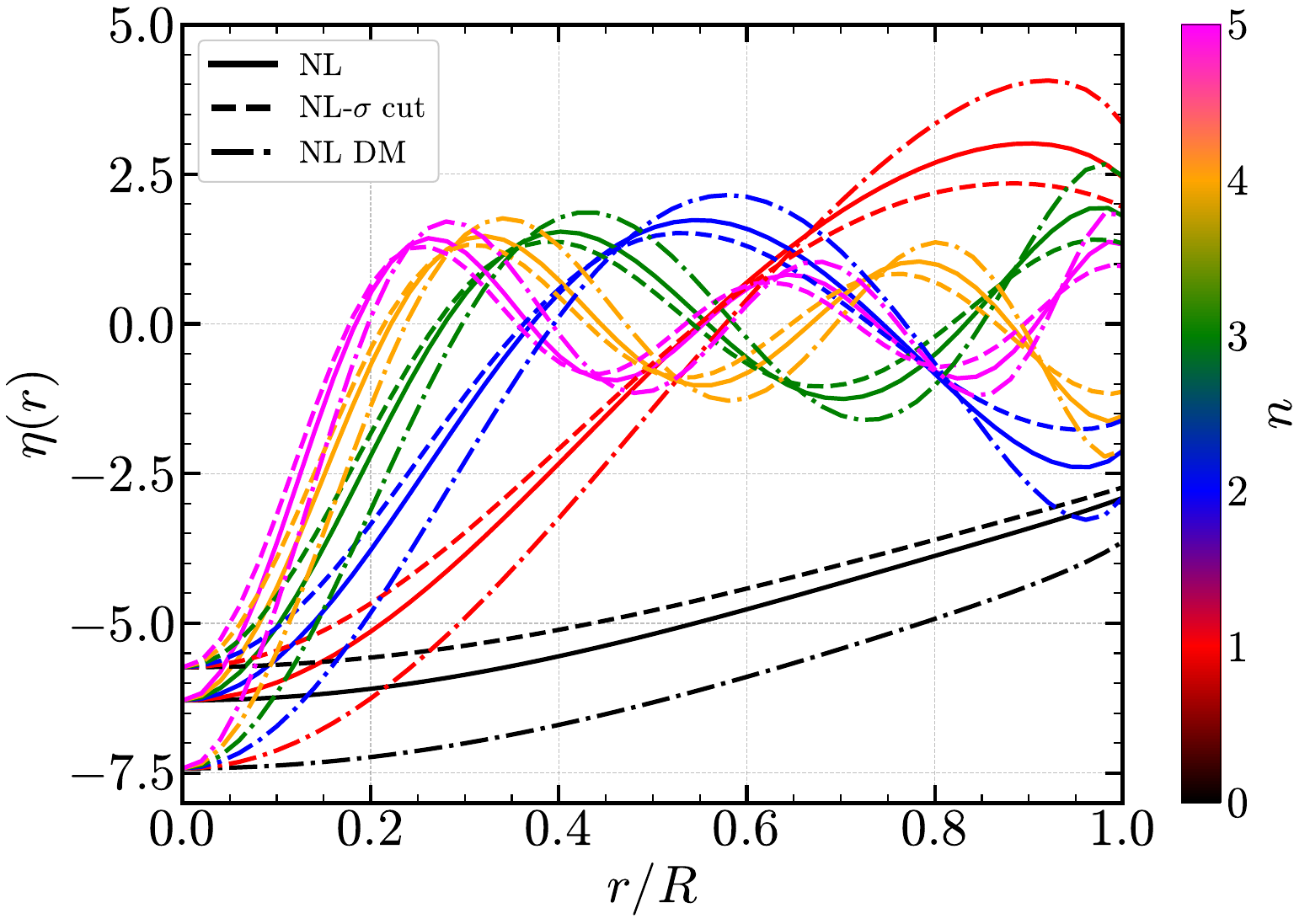}
			 	\end{minipage}
		 		\begin{minipage}[t]{0.49\textwidth}
			 		\includegraphics[width=\textwidth]{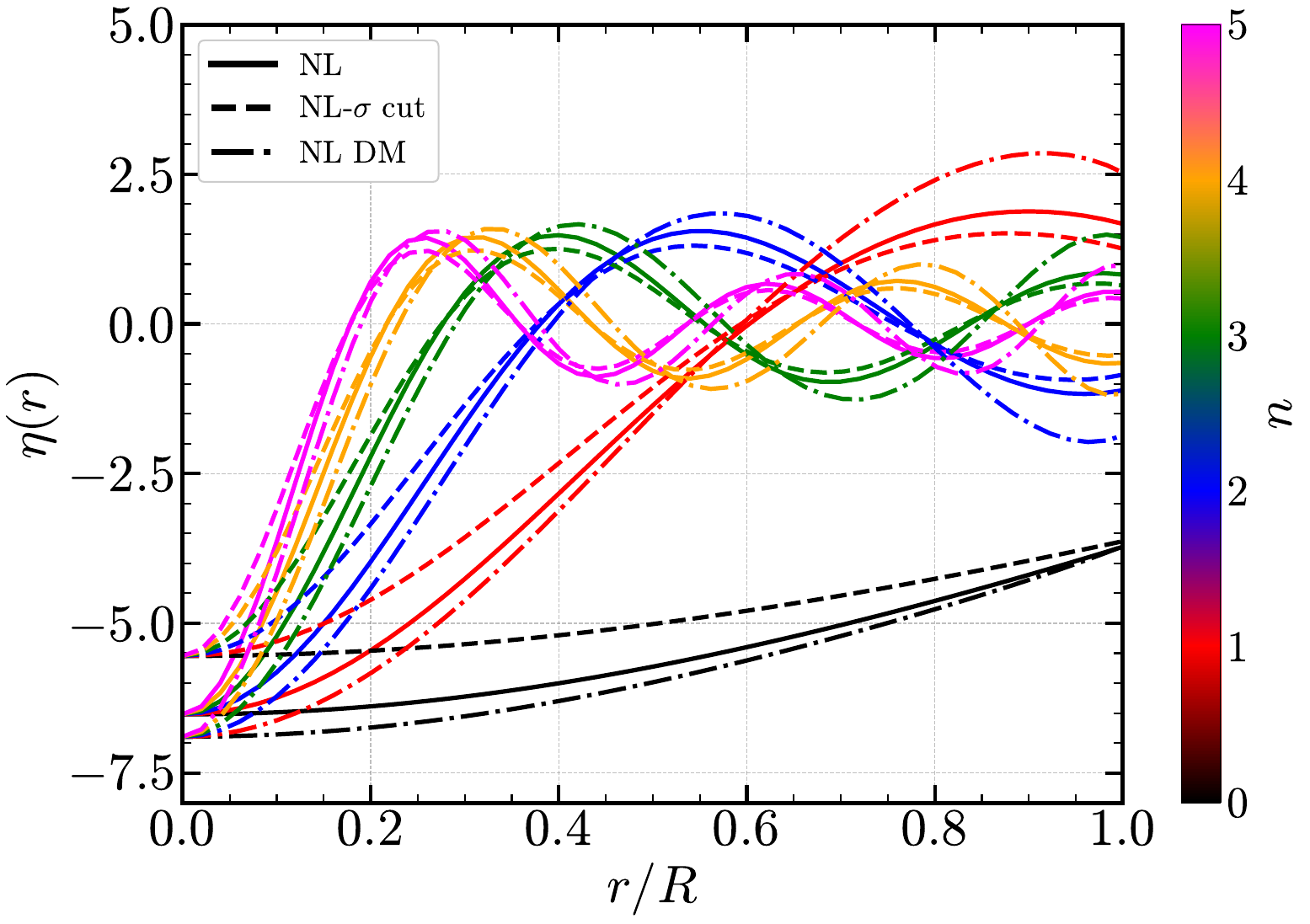}
			 	\end{minipage}
			 			\caption{Left: The radial pressure perturbation ${\eta(r)}$ = ${\Delta r/r}$ as a function of dimensionless radius distance ${r/R}$ The colorbar represents the modes: ${f}$-mode ($n = 0$) and lower-order ${p}$-modes ($n = 1-5$) for soft NL, NL-$\sigma$ cut, and NL DM EoS. Right: The same as the left plot, but for stiff EoSs. All modes are calculated at 1.4\,${M_{\odot}}$. }
		\label{fig:eta}	 	
     \end{figure*}
     
Solving the Sturm-Liouville eigenvalue equations for $\omega$, we calculate the first 10 modes $\omega_n$, with $n = 0$ corresponding to the $f$-mode and $n = 1-9$ as high-order $p$-modes. The radial displacement perturbation profile $\xi(r)$ as a function of the dimensionless radial coordinate $r/R$ is shown in Fig.~\ref{fig:ksi}. The left (right) panel presents the profile for soft (stiff) NL, NL-$\sigma$ cut, and NL DM EoS. We display only the fundamental $f$-mode ($n = 0$) and the first few lower-order $p$-modes ($n = 1-5$) in the plot, with the colorbar representing the modes. The solid, dashed, and dot-dashed lines correspond to the NL, NL-$\sigma$ cut, and NL DM EoS, respectively.  In the region $0<r<R$, the $n$th mode exhibits exactly $n$ nodes in the $\xi(r)$ profile, following the characteristics of a Sturm-Liouville system. The amplitude of $\xi_n(r)$ for each frequency mode $\nu_n$ is largest near the center and diminishes toward the surface. Lower-order modes exhibit a smooth decline, whereas higher-order modes display small oscillations, which become more pronounced for even higher modes.  

In the left panel, corresponding to the soft EoS, the fundamental mode $\xi_0$ for NL and NL-$\sigma$ cut exhibits a decrease in amplitude near the surface, with the effect being more pronounced for the NL-$\sigma$ cut EoS. The inclusion of dark matter in the NL DM EoS further softens the EoS, leading to a larger amplitude compared to the other cases. For higher modes, the NL DM EoS consistently exhibits the largest amplitude up to a point where all three lines intersect each other. After this point, its behavior changes as it approaches the surface of the star.  

In the right panel, which corresponds to the stiff EoS, the NL-$\sigma$ cut EoS displays a larger amplitude for the $f$-mode than the NL DM case. Additionally, the overall amplitude for stiff EoSs is slightly larger than that for soft EoSs. However, the higher-order modes behave similarly in both cases, with a marginally larger amplitude near the stellar surface in the stiff EoS.  

The radial pressure perturbation profile, $\eta(r)$, for the same EoSs, is shown in Fig.~\ref{fig:eta} as a function of the dimensionless radius $r/R$. The amplitude $\eta_n(r)$ is largest near both the center and the surface of the star.  

For the soft EoS (left panel), the amplitude of $\eta_0(r)$ for the NL DM case is lower than that for NL and NL-$\sigma$ cut at both the center and the surface. However, for higher modes, the amplitude remains small at the center but increases toward the surface. In the stiff EoS case (right panel), the overall amplitude remains lower near the stellar surface.  So, while the $\xi(r)$ modes for EoSs with different compositions follow a similar trend with a small difference in the amplitude for the DM, NL DM EoS, the $\eta(r)$ amplitudes ($f$ and low $p$-modes) show a clear distinction between different compositions considered in this work, for both soft and stiff EoSs.

Since the oscillations $\eta(r)$ are directly proportional to the Lagrangian pressure variation $\Delta P$, the amplitudes of consecutive modes, $\eta_{n+1}(r)$ and $\eta_n(r)$, are significant near the surface. However, due to their opposite signs, their contributions cancel out, ensuring that the boundary condition $P(r=R) = 0$ is satisfied. Consequently, the differences $\eta_{n+1} - \eta_n$ and $\xi_{n+1} - \xi_n$ are more sensitive to the stellar core. This sensitivity suggests that the frequency separation $\Delta \nu_n = \nu_{n+1} - \nu_n$ serves as an observational signature of the star’s innermost layers.

It is noteworthy that $\Delta \nu_n$ exhibits a dependence on $\nu_n$, providing evidence that the microphysics, or equivalently the EoS of a NS interior is imprinted on the large frequency separation. This characteristic, which is also well established in main sequence stars such as the Sun, arises when $\Delta \nu_n$ becomes constant and is proportional to $\sqrt{M/R^3}$, a quantity independent of $\nu_n$. Superimposed on this constant value, any structural discontinuity or glitch in the star produces a small oscillatory modulation in $\Delta \nu_n$. In the context of NSs, the amplitude of this oscillation is proportional to the magnitude of the discontinuity, which originates from the rapid variation of the speed of sound or the relativistic adiabatic index across the transition layer separating the inner and outer core (see Eq.~\ref{gamma}). Moreover, the period of the oscillatory signature is linked to the depth at which the discontinuity occurs beneath the star’s surface.

\begin{figure}[t]	 		
  \includegraphics[width=0.47\textwidth]{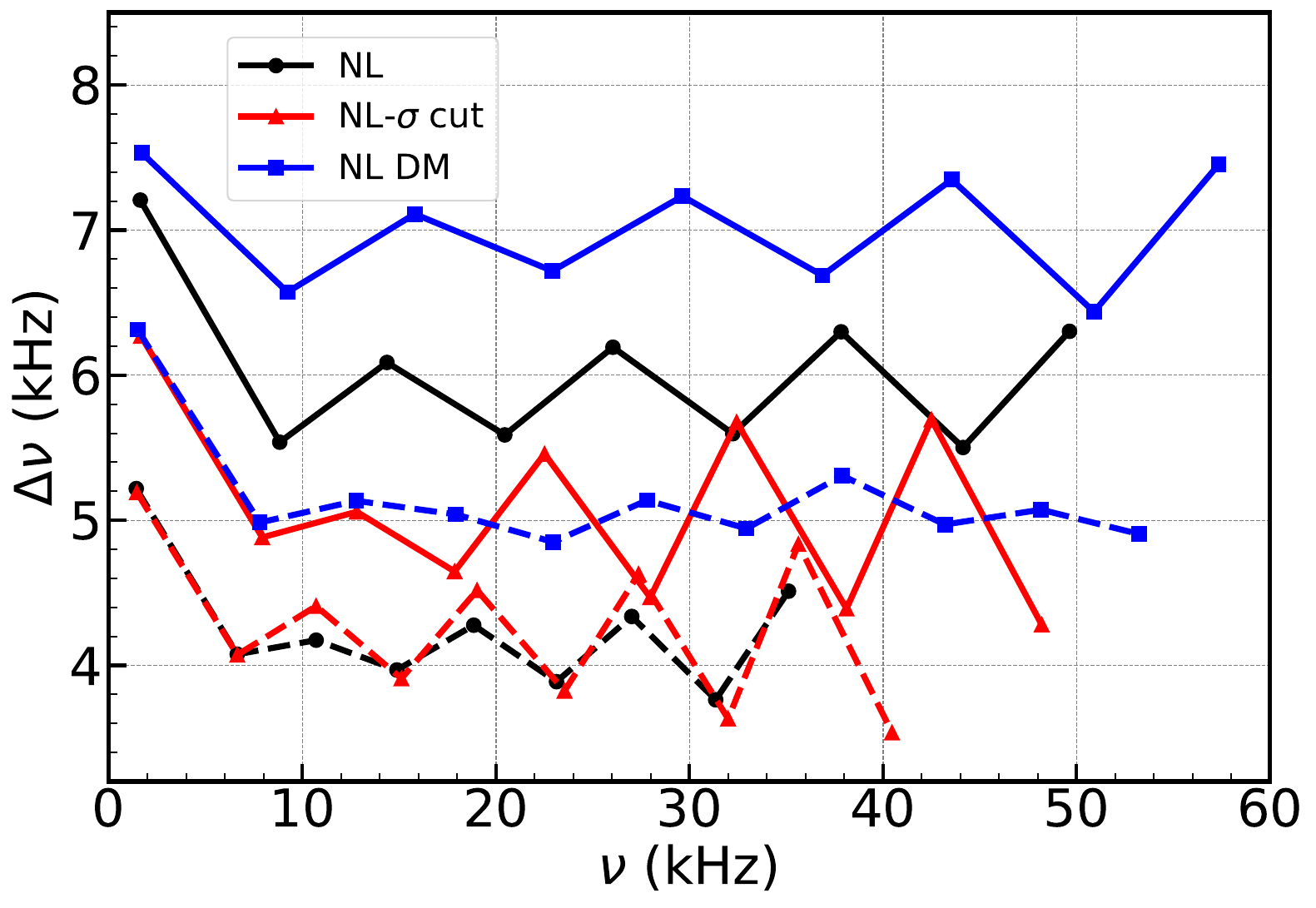}
			 			\caption{Frequency difference ${\Delta \nu_n}$ = ${\nu_{n+1}}$ - ${\nu_n}$ versus ${\nu_n}$ in kHz for NL, NL-$\sigma$ cut, and NL DM EoS. The solid (dashed) lines correspond to the soft (stiff) EoSs. }
		\label{fig:freq}	 	
     \end{figure}

 Fig.~\ref{fig:freq} illustrates the frequency separation, $\Delta \nu$ (in kHz), as a function of the oscillation frequency, $\nu$ (in kHz), for NS matter modeled by three different EoSs: NL, NL-$\sigma$ cut, and NL DM. The frequencies for the fundamental mode ($n = 0$) and the higher $p$-modes ($n = 1-9$) are displayed in Table~\ref{table1}. In this figure, the solid lines represent softer EoSs, while the dashed lines correspond to stiffer ones. The variations observed in $\Delta \nu$ as $\nu$ increases can be primarily attributed to the transition between the crust and the core of the neutron star, as discussed in Ref.~\cite{Rather:2023dom,  Rather:2024hmo, Sen:2022kva, Routaray:2022utr}. In the crust, where the density is relatively low and nuclei coexist with free neutrons, the oscillation modes are influenced by the inhomogeneous composition. As the density exceeds a critical threshold, the matter becomes uniform in the core, and this crust–core boundary gives rise to mode coupling or conversion effects that result in fluctuations in the oscillation frequencies. Softer EoSs, which produce steeper density gradients and more compact stars, tend to enhance the interaction between crustal and global modes, leading to more pronounced fluctuations in $\Delta \nu$. Conversely, stiffer EoSs generate smoother transitions in density and support larger masses and radii, shifting the oscillation modes and typically shortening or decreasing the frequency separation.

\begin{table}[ht]
\centering
		\caption{10 lowest-order radial oscillation frequencies, ${\nu}$ in (kHz) for different compositions. For each EoS (soft and stiff), the frequencies are calculated at 1.4\,${M_{\odot}}$.\label{table1} }
\begin{tabular}{ p{1.5cm}p{1.5cm}p{1.5cm}p{1.5cm} }
 \hline
\multirow{2}{*}{Order (n)} &\multicolumn{3}{c}{EoS} \\
 \cline{2-4}
  & NL & NL-$\sigma$ cut & NL DM \\
 \hline
 & &  soft &  \\
\hline 
0   &1.62&1.65&1.69\\
1	&8.82&7.92&9.02\\
2	&14.36&12.80&15.80\\
3	&20.45&17.86&22.91\\
4	&26.054&22.51&29.63\\
5	&32.23&27.97&36.76\\
6	&37.83&32.43&43.55\\
7	&44.13&38.11&50.70\\
8	&49.63&42.50&57.33\\
9	&55.93&48.20&64.51\\
\hline 
& &  stiff &  \\
\hline 
0   &1.41 &1.44 &1.48\\
1	&6.63 &6.63 &7.79\\ 
2	&10.71 &10.70 &12.78\\
3	&14.88 &15.11 &17.71\\
4	&18.85 &19.02 &22.65\\
5	&23.12 &23.54 &27.60\\
6	&27.01 &27.36 &32.56\\
7	&31.35 &31.99 &37.52\\
8	&35.11 &35.63 &42.49\\
9	&39.63 &40.46 &47.46\\
 \hline
\end{tabular}
\end{table}

For the soft NL EoS, the frequency separation for higher modes fluctuates within a range of roughly 0.5~kHz, whereas in the NL-$\sigma$ cut and NL DM cases, the separation is slightly larger. The modification introduced by the NL-$\sigma$ cut potential affects the density dependence of the nuclear interaction, rendering the EoS more compressible and thereby producing a moderate increase in $\Delta \nu_n$. Similarly, the inclusion of dark matter in the NL DM model further softens the EoS, leading to an enhanced frequency separation relative to the pure NL case. The fluctuations for higher modes are more pronounced in the NL-$\sigma$ cut and NL DM scenarios compared to the standard NL EoS.

In the stiff EoS regime, the overall frequency separation and fluctuation between different modes is generally smaller. Among them, the NL DM EoS displays almost a constant $\Delta \nu$, approaching a value of approximately 5.0~kHz, while both the NL and NL-$\sigma$ cut models exhibit a decrease in the separation, but with large fluctuations compared to the DM EoS. This decrease is a direct consequence of the stiffening effects induced by the NL and NL-$\sigma$ cut, which reduces the restoring force of the stellar perturbations.

The effect of dark matter on the star's internal structure is clearly seen in the NL DM model. The presence of dark matter alters both the mass distribution and the pressure balance within the star, leading to shifts in the oscillation modes. These shifts can either accentuate or dampen the observed fluctuations in $\Delta \nu$ when compared to the NL and NL-$\sigma$ cut cases.

The overall trends observed in Fig.~\ref{fig:freq} are in good agreement with the theoretical predictions of NS oscillations as influenced by the EoS. \citet{Lattimer:2000nx} provides an extensive review of how variations in the EoS stiffness affect the mass-radius relationship and the stellar oscillation modes, while \citet{Glendenning:1997wn} and \citet{haensel2006neutron} emphasize the crucial role of crust–core transitions in determining both the nuclear composition and the dynamical behavior of compact stars. Ref.~\cite{Routaray:2022utr} also studied the frequency difference at different DM momenta, with and without the crust part, showing how the oscillatory behavior changes when the crust part is not taken into account. 

Our study of radial oscillations with DM differs from others \cite{Sen:2022kva, PhysRevD.101.063025, PhysRevD.98.083001, Routaray:2022utr} in the sense that we use a neutron decay anomaly model, which provides us with a scenario where a chemical equilibrium is established between ordinary matter and the
dark sector. Furthermore, we study the oscillation modes considering two extreme parameter values, called soft and stiff, that are well within the astrophysical constraints. This gives us a detailed description of how the modes, especially $f$ and $p$-modes, change and, more importantly, how the large frequency separation can help us distinguish between EoSs with different compositions and different parameter values.

\begin{figure}[h]	 		
  \includegraphics[width=0.47\textwidth]{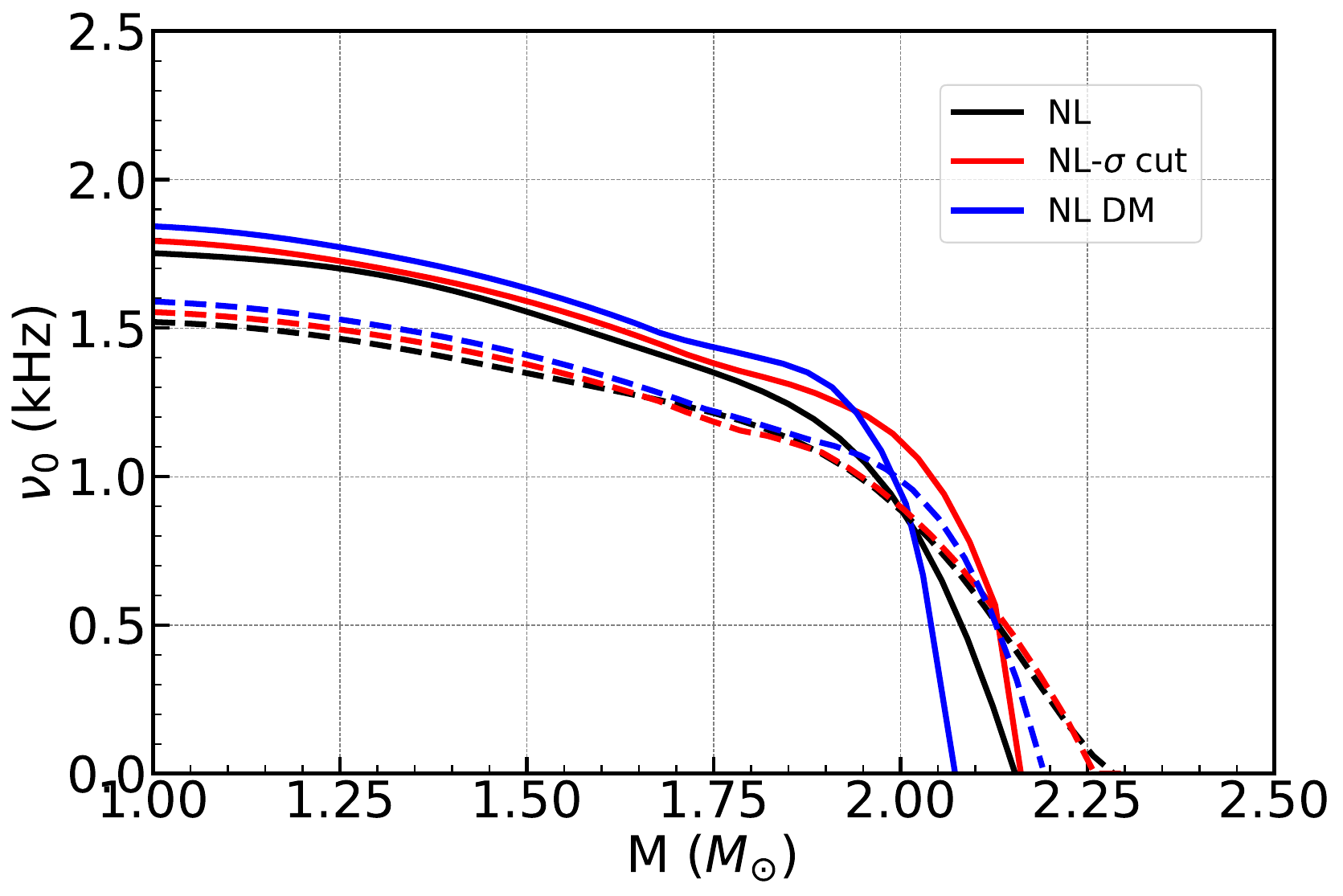}
			 			\caption{Radial $f$-mode ($n = 0$) frequency as a function of mass for NL, NL-$\sigma$ cut, and NL DM EoS. The solid (dashed) lines correspond to the soft (stiff) EoSs. }
		\label{fig:fM}	 	
     \end{figure}

   Fig.~\ref{fig:fM} shows the variation of fundamental $f$-mode ($n = 0$) frequency as a function of stellar mass for all the EoS studied in this work. This study focuses on a comprehensive examination of the connection between radial oscillation and the stability of NS. This figure shows how the frequency decreases with the increasing mass and how the $f$-mode oscillates at higher frequencies for softer EoS than the stiff ones.
As shown in earlier studies \cite{Sun:2021cez, Routaray:2022utr, Rather:2024hmo}, the $f$-mode swiftly nears zero, where the maximum numerical NS mass is attained along the MR curve. This aligns with the stability condition $\partial{M}/\partial{\rho_c} >$ 0. 

\subsection{Relevance for Current and Future GW Detectors}

We investigate the detectability of gravitational waves (GWs) emitted from both the fundamental ($f$) and first pressure ($p_1$) modes of NS oscillations. Following the formalism established by \cite{Thakur:2024ijp,VasquezFlores:2018tjl}, the energy radiated via GWs from these modes can be expressed as:

\begin{equation}
\begin{split}
\frac{E_{\text{GW}}}{M_\odot c^2} = 3.471 \times 10^{36} 
\left( \frac{S}{N} \right)^2 
\left( \frac{1 + 4Q^2}{Q^2} \right)
\left( \frac{D}{10\,\text{kpc}} \right)^2 \\
\times 
\left( \frac{f}{1\,\text{kHz}} \right)^2 
\left( \frac{S_n}{1\,\text{Hz}^{-1}} \right)
\end{split}
\label{E_gw}
\end{equation}
We estimate the minimum energy that must be emitted in order to achieve a signal-to-noise ratio (S/N) greater than $5$. In Eq.~\ref{E_gw}, the quality factor is given by $ Q = \pi f \tau $, where $ \tau $ is the damping time, $ f $ is the oscillation frequency of either the $ f $-mode or $ p_1 $-mode, and $ S_n $ represents the spectral noise density. We compute the gravitational wave energy $ E_{\text{GW}} $ for both the canonical and maximum mass configurations for the $ f $-mode.

We examine two categories of gravitational wave detectors. The first, with a sensitivity of approximately $ S_n^{1/2} \sim 2 \times 10^{-23}\, \mathrm{Hz}^{-1/2} $, represents the typical performance of Advanced LIGO and Virgo in the kHz frequency range\,\cite{LIGOScientific:2017vwq}. The second reflects the projected sensitivity of future third-generation detectors such as the Einstein Telescope, expected to reach around $ S_n^{1/2} \sim 10^{-24}\, \mathrm{Hz}^{-1/2} $ in a similar frequency band\,\cite{LIGOScientific:2016wof}. Regarding source distances, we consider two representative cases: one with a NS located in the Virgo cluster at a distance of roughly $ D \sim 15\, \mathrm{Mpc} $, and another situated within our Galaxy at about $ D \sim 10\, \mathrm{kpc} $.

The first part of Table~\ref{E_GW} presents estimates of the gravitational wave energy $E_{\text{GW}}$ emitted in the fundamental $f$-mode by NSs with a mass of $1.4\,M_\odot$, for the models considered in this work. 

The $f$-mode frequency lies between $1.73$ and $1.85$~kHz, with damping times ranging from $0.20$ to $0.24$~s, showing minimal variation among models. At a distance of 10~kpc, the estimated energy required for detection is approximately $1.05 \times 10^{-7}\,M_\odot c^2$, which is significantly lower than the typical energy released in a supernova ($10^{-5}$–$10^{-6}\,M_\odot c^2$), making detection at this distance feasible. However, at $15$~Mpc, the required energy increases sharply to about $0.23$–$0.26\,M_\odot c^2$, which far exceeds expected emission levels, thereby rendering such distant signals undetectable with current sensitivity levels. At a spectral noise density of $S_n = 1 \times 10^{-24}~\mathrm{Hz}^{-1}$, the estimated energy required for detecting gravitational waves is approximately $10^{-10}\,M_\odot c^2$ at $10$~kpc, while at $15$~Mpc, it ranges from $5.91 \times 10^{-4}$ to $6.60 \times 10^{-4}\,M_\odot c^2$.

We further analyze how the microphysical modifications in the EoS, specifically through the $\sigma$-cut potential and dark matter, impact the gravitational wave energy emitted by NSs. In the NL-$\sigma$ cut model, the stiffening at intermediate densities combined with suppressed scalar interactions at high densities leads to lower stellar compactness and longer damping times. This results in a reduction of the total GW energy radiated by $f$-mode oscillations. On the other hand, the inclusion of DM via the neutron decay anomaly model softens the EoS, producing more compact stars with higher mode frequencies and shorter damping times. These changes enhance the gravitational wave luminosity and shift the $f$-mode signal toward more detectable regimes. Thus, the $\sigma$-cut and DM components introduce clearly distinguishable trends in the GW energy output, highlighting their relevance for future asteroseismic measurements with advanced detector.

\begin{table*}[htbp]
\centering
\caption{Gravitational wave energy estimates of the $f$-mode frequency for NL, NL-$\sigma$ cut, and NL DM models at $D = 10$ kpc and $15$ Mpc. The corresponding frequency, $f_{\text{mode}}$ (in kHz), and the damping time, $\tau$ (in sec), are shown for two configurations: $1.4\,M_{\odot}$ and maximum mass.}
\begin{tabular}{|c|c|c|c|c|c|}
\hline
\multicolumn{6}{|c|}{$1.4\,M_{\odot}$ Configuration} \\
\hline
Distance & Model & $f_{\text{mode}}$ (kHz) & $\tau$ (s) & $E_{\text{GW}}/(M_\odot c^2)$  & $E_{\text{GW}}/(M_\odot c^2)$ \\
&&&& ($S_n = 2 \times 10^{-23}~\mathrm{Hz}^{-1}$) &  ($S_n = 1 \times 10^{-24}~\mathrm{Hz}^{-1}$) \\
\hline
\multirow{6}{*}{10 kpc}
& NL soft             & 1.81 & 0.217 & $1.13 \times 10^{-7}$ & $2.84 \times 10^{-10}$ \\
& NL stiff            & 1.74 & 0.235 & $1.05 \times 10^{-7}$ & $2.62 \times 10^{-10}$ \\
& NL-$\sigma$ cut soft    & 1.83 & 0.212 & $1.16 \times 10^{-7}$ & $2.90 \times 10^{-10}$ \\
& NL-$\sigma$ cut stiff   & 1.73 & 0.240 & $1.03 \times 10^{-7}$ & $2.59 \times 10^{-10}$ \\
& NL DM soft          & 1.85 & 0.209 & $1.18 \times 10^{-7}$ & $2.96 \times 10^{-10}$ \\
& NL DM stiff         & 1.80 & 0.219 & $1.12 \times 10^{-7}$ & $2.81 \times 10^{-10}$ \\
\hline
\multirow{6}{*}{15 Mpc}
& NL soft             & 1.81 & 0.217 & $0.25$ & $6.39 \times 10^{-4}$ \\
& NL stiff            & 1.74 & 0.235 & $0.23$ & $5.91 \times 10^{-4}$ \\
& NL-$\sigma$ cut soft    & 1.83 & 0.212 & $0.26$ & $6.53 \times 10^{-4}$ \\
& NL-$\sigma$ cut stiff   & 1.73 & 0.242 & $0.23$ & $5.80 \times 10^{-4}$ \\
& NL DM soft          & 1.85 & 0.209 & $0.26$ & $6.60 \times 10^{-4}$ \\
& NL DM stiff         & 1.80 & 0.219 & $0.25$ & $6.30 \times 10^{-4}$ \\
\hline
\hline 
\multicolumn{6}{|c|}{Maximum Mass Configuration} \\
\hline
\hline 
Distance & Model & $f_{\text{mode}}$ (kHz) & $\tau$ (s) & $E_{\text{GW}}/(M_\odot c^2)$  & $E_{\text{GW}}/(M_\odot c^2)$ \\
&&&& ($S_n = 2 \times 10^{-23}~\mathrm{Hz}^{-1}$) &  ($S_n = 1 \times 10^{-24}~\mathrm{Hz}^{-1}$) \\
\hline
\multirow{6}{*}{10 kpc}
& NL soft             & 2.34 & 0.141 & $1.90 \times 10^{-7}$ & $4.75 \times 10^{-10}$ \\
& NL stiff            & 2.26 & 0.153 & $1.70 \times 10^{-7}$ & $4.43 \times 10^{-10}$ \\
& NL-$\sigma$ cut soft    & 2.29 & 0.139 & $1.82 \times 10^{-7}$ & $4.55 \times 10^{-10}$ \\
& NL-$\sigma$ cut stiff   & 2.17 & 0.148 & $1.63 \times 10^{-7}$ & $4.08 \times 10^{-10}$ \\
& NL DM soft          & 2.43 & 0.136 & $2.04 \times 10^{-7}$ & $5.12 \times 10^{-10}$ \\
& NL DM stiff         & 2.33 & 0.145 & $1.88 \times 10^{-7}$ & $4.71 \times 10^{-10}$ \\
\hline
\multirow{6}{*}{15 Mpc}
& NL soft             & 2.34 & 0.141 & $0.42$ & $1.06 \times 10^{-3}$ \\
& NL stiff            & 2.26 & 0.153 & $0.39$ & $9.97 \times 10^{-4}$ \\
& NL-$\sigma$ cut soft    & 2.29 & 0.139 & $0.40$ & $1.02 \times 10^{-3}$ \\
& NL-$\sigma$ cut stiff   & 2.17 & 0.148 & $0.36$ & $9.19 \times 10^{-4}$ \\
& NL DM soft          & 2.43 & 0.136 & $0.46$ & $1.15 \times 10^{-3}$ \\
& NL DM stiff         & 2.33 & 0.145 & $0.42$ & $1.06 \times 10^{-3}$ \\
\hline
\end{tabular}
\label{E_GW}
\end{table*}

For the maximum mass configuration, which are shown in the lower panel of Table \ref{E_GW}, frequencies are higher (2.17–2.43 kHz) with shorter damping times (0.14–0.15 s), leading to thresholds of roughly $1.63\times10^{-7}$–$2.04\times10^{-7}\,M_\odot c^2$ at 10 kpc (with $S_n = 2\times10^{-23}$) and $0.36$–$0.46\,M_\odot c^2$ at 15 Mpc, which drop to $\sim4\times10^{-10}$–$5\times10^{-10}\,M_\odot c^2$ and $\sim9\times10^{-4}$–$1.15\times10^{-3}\,M_\odot c^2$ with $S_n = 1\times10^{-24}$. In short, nearby sources (10 kpc) are detectable with much lower energy thresholds, while even with improved sensitivity, distant sources (15 Mpc) require unrealistically high energies.
 
For a $1.4\,M_\odot$ NS, the GW energy emitted in the $p_1$-mode lies in the range of $1.19 \times 10^{-6}$ to $1.47 \times 10^{-6}\,M_\odot c^2$ when observed at a distance of $10~\mathrm{kpc}$, assuming a spectral noise density of $S_n = 2 \times 10^{-23}~\mathrm{Hz}^{-1}$. At a distance of $15~\mathrm{Mpc}$, the corresponding energy lies between $2.69 \times 10^{-3}$ and $3.32 \times 10^{-3}\,M_\odot c^2$.

For a more sensitive detector such as the Einstein Telescope, with $S_n = 1 \times 10^{-24}~\mathrm{Hz}^{-1}$, the detectable energy at $10~\mathrm{kpc}$ lies in the range $2.99 \times 10^{-9}$ to $3.69 \times 10^{-9}\,M_\odot c^2$, and at $15~\mathrm{Mpc}$ it lies between $6.73 \times 10^{-3}$ and $8.30 \times 10^{-3}\,M_\odot c^2$.

Therefore, gravitational waves from $p_1$-mode oscillations of NSs located within our galaxy (e.g., at $10~\mathrm{kpc}$) are potentially detectable by current detectors such as Advanced LIGO and VIRGO, and more certainly by third-generation detectors like the Einstein Telescope. However, for extra-galactic sources—such as those in the Vela Cluster at $\sim15~\mathrm{Mpc}$—the GW signals are too weak to be detected with existing detectors and are only marginally within the reach of future observatories like the Einstein Telescope \cite{Abac:2025saz}. Our results are in agreement with previous works~\cite{Thakur:2024ijp,VasquezFlores:2018tjl}

\section{Discussion and Outlook}
\label{summary}
In this work, we have investigated the impact of both the $\sigma$-cut potential and dark matter on the oscillation modes of neutron stars within the relativistic mean-field framework. Three different equations of state (EoSs) were analyzed: the standard nonlinear (NL) model, the NL model with a modified $\sigma$-cut potential (NL-$\sigma$ cut), and the NL model incorporating dark matter (NL DM). Both soft and stiff parameterizations were explored to cover the range of astrophysically allowed behaviors.

Our study shows that the introduction of soft parameterizations for the $\sigma$-cut potential enhances the stiffness of the EoS, resulting in a larger radius across the density range. For the stiff parameterization, the effect is soft at low to intermediate densities, leading to a smaller radius than the standard NL parameterization,  while its effect at very high densities ($M \approx 1.5\,M_{\odot}$) leads to a noticeable increase in the radius. In contrast, the inclusion of dark matter consistently softens the EoS across the density range, resulting in distinct modifications in the mass-radius relationship. These changes are directly reflected in the oscillation spectra:
\begin{itemize}
    \item Non-radial oscillations: The $f$-mode frequencies and damping times are sensitive to the compactness and mean density of the neutron star. Our calculations indicate that dark matter admixture can lead to higher $f$-mode frequencies at the canonical mass of $1.4\,M_{\odot}$, while the damping times exhibit systematic variations with the EoS stiffness. Our $p_1$-mode analysis ($\ell = 2$) across the NL, NL-$\sigma$ cut, and NL DM models reveals distinctive oscillation characteristics. Frequencies range from 5-6.5 kHz, generally increasing with stellar mass before slightly decreasing at higher masses, with NL DM models exhibiting the highest frequencies. Damping times follow an inverse pattern, with NL-$\sigma$ stiff showing the longest time (10.176 s) and NL soft the shortest (5.851 s). Dark matter softens the EoS, creating more compact stars with higher frequencies but shorter damping times, while modified sigma potentials increase rigidity, reducing frequencies but extending damping times.
    
    \item Radial oscillations: Our analysis of neutron star oscillation modes across NL, NL-$\sigma$ cut, and NL DM models reveals distinctive eigenmode characteristics. The radial displacement perturbation $\xi(r)$ shows highest amplitudes near the center, with NL DM exhibiting larger values in soft EoS models due to its softer nature. Pressure perturbation profiles $\eta(r)$ display clear composition-dependent differences with amplitudes peaking at both center and surface. The frequency separation $\Delta\nu_n$ serves as a sensitive probe of stellar interiors, with soft NL EoS showing fluctuations within $\sim$0.5 kHz for higher modes, while NL-$\sigma$ cut and NL DM demonstrate larger separations. In stiff EoSs, NL DM maintains nearly constant $\Delta\nu$ ($\sim$5.0 kHz), while other models show decreasing separation with larger fluctuations. The amount of DM present inside the star, even a small fraction, has a strong impact on the large separation, which, depending on the order of the mode $n$, is sensitive to different regions of the star.  These variations reflect crust-core transitions, with softer EoSs enhancing crustal-global mode interactions. Our neutron decay anomaly model establishes chemical equilibrium between ordinary and dark matter, offering unique insights into how dark matter alters oscillation characteristics. This approach demonstrates how frequency separation can effectively distinguish between different EoS compositions and parameters.
    
\end{itemize}
Furthermore, the quasi-universal relations established for the $f$-mode frequency and damping time are consistent with previous studies, yet they reveal noticeable modifications due to the effects of the $\sigma$-cut and dark matter.
 Notably, for $f$-mode frequency and the mean stellar density, our previously established fits IR-1 and IR-2, derived from hyperons and $\Delta$ baryons without and with phase transitions, respectively, perfectly describe our current data that includes dark matter. This remarkable consistency across diverse matter compositions demonstrates the robust nature of these quasi-universal relations, outperforming other parameterizations in their predictive power. Such universality provides a powerful framework for interpreting observational data without requiring detailed knowledge of specific matter constituents. These findings underline the potential of asteroseismology as a tool for probing the dense matter equation of state and the possible presence of exotic components in NSs.

We have estimated the minimum gravitational wave energy required to achieve a signal-to-noise ratio greater than 5 for the $f$- and $p_1$-mode oscillations in neutron stars. Our study considered both current-generation detectors, such as Advanced LIGO and Virgo, and future third-generation observatories like the Einstein Telescope, across two representative source distances: a galactic neutron star at $10~\mathrm{kpc}$ and an extra-galactic one in the Vela Cluster, located at approximately $15~\mathrm{Mpc}$. The results show that galactic sources are well within the detection capability of existing detectors, with energy thresholds significantly lower than the energy released in typical supernovae. In contrast, detecting sources in the Vela Cluster requires gravitational wave energies that are orders of magnitude higher than expected from realistic neutron star oscillations. Even with the improved sensitivity of future detectors, such distant signals remain largely inaccessible, especially for the $p_1$-mode. Therefore, gravitational wave signals from neutron star oscillations are promising candidates for detection from within our galaxy, but are energetically out of reach for extra-galactic sources like those in the Vela Cluster.

The present work highlights several avenues for future research:
\begin{enumerate}
    \item Refinement of the EoS: Further improvements in the microphysical modeling of the $\sigma$-cut potential and dark matter interactions are needed. A more detailed treatment of these effects could lead to a better understanding of the transition between low- and high-density regimes.
    \item Extended Mode Analysis: In addition to the $f$-mode, it would be beneficial to study other non-radial modes (e.g., $g$ and $r$-modes) and assess their potential as diagnostic tools for the internal structure of neutron stars. In particular, $g$-modes are highly sensitive to the internal composition and stratification of neutron stars, making them valuable probes for detecting composition gradients and phase transitions within the stellar core.
    \item Observational Constraints: With the advent of next-generation gravitational wave detectors and multimessenger observations, future work would aim to incorporate observational constraints to narrow down the model parameter space.
    \item Numerical Simulations: Studying the oscillation modes for a finite temperature EoS with $\sigma$-cut and DM along with the advancing numerical simulations to include the effects of rotation and magnetic fields will be crucial, as these factors are expected to influence the oscillation spectra and overall stability of neutron stars.
\end{enumerate}

Addressing these aspects will not only refine our theoretical models but also enhance the prospects of using NS asteroseismology as a powerful probe of dense matter physics. An important direction for future work involves extending the study of both radial and non-radial oscillation modes in the presence of exotic phases of matter, such as color-superconducting quark matter \cite{Gholami:2024ety}, and in more complex stellar structures like hybrid stars and twin star configurations \cite{Christian:2025dhe}. These studies could provide deeper insights into the role of phase transitions in determining the oscillation spectra, and may ultimately offer observational signatures that help distinguish between different models of the dense matter EoS.

\section{Acknowledgement}
I.A.R. acknowledges support from the Alexander von Humboldt Foundation. 
Y.~Lim is supported by the National
Research Foundation of Korea(NRF) grant funded by the
Korea government(MSIT)(No. RS-2024-00457037) and
by Global - Learning \& Academic research institution for Master's·PhD students, 
and Postdocs(LAMP) Program of the National Research Foundation of Korea(NRF) grant funded by the Ministry of Education(No.  RS-2024-00442483).
Y. Lim is also supported by the Yonsei University Research Fund of 2024-22-0121. 

%


\end{document}